\documentclass{article}
\usepackage{epsfig}
\usepackage{cite}
\usepackage{float}
\usepackage{mathrsfs}
\usepackage{amsfonts}
\usepackage{array}
\usepackage{epsfig}
\usepackage{amsmath}    
\usepackage{amssymb}   
\usepackage{graphicx}   
\usepackage{verbatim}   
\usepackage{color}      
\usepackage{slashed}

\usepackage{listings}
\usepackage{amsmath,amsfonts,amssymb,epsfig,comment,xspace,listings,cite}
\usepackage{graphicx,caption,subcaption}

\definecolor{maroon}{cmyk}{0, 0.87, 0.68, 0.32}
\definecolor{halfgray}{gray}{0.55}
\definecolor{slha_frame}{RGB}{207, 207, 207}
\definecolor{slha_bg}{RGB}{247, 247, 247}
\definecolor{slha_red}{RGB}{186, 33, 33}
\definecolor{slha_green}{RGB}{0, 128, 0}
\definecolor{slha_cyan}{RGB}{64, 128, 128}
\definecolor{slha_purple}{RGB}{170, 34, 255}

\definecolor{mathematica_frame}{RGB}{207, 207, 207}
\definecolor{mathematica_bg}{RGB}{247, 247, 247}
\definecolor{mathematica_red}{RGB}{186, 33, 33}
\definecolor{mathematica_green}{RGB}{0, 128, 0}
\definecolor{mathematica_cyan}{RGB}{64, 128, 128}
\definecolor{mathematica_purple}{RGB}{170, 34, 255}

\lstnewenvironment{MIN}[1][]{%
  \renewcommand{\thelstnumber}{In[\arabic{lstnumber}]}
  \lstset{language=MathIn,numbers=left,basicstyle=\ttfamily,#1}%
}{%
}

\lstnewenvironment{MOUT}[1][]{%
  \renewcommand{\thelstnumber}{Out[\arabic{lstnumber}]}
  \lstset{language=MathOut,numbers=left,basicstyle=\ttfamily,#1}%
}{%
}

\usepackage{listings}
\lstdefinelanguage{SLHA}{
    morekeywords={block,Block,BLOCK,decay,Decay,DECAY},%
    sensitive=true,%
    morecomment=[l]\#,%
    morestring=[b]',%
    morestring=[b]",%
    morestring=[s]{'''}{'''},
    morestring=[s]{"""}{"""},
    morestring=[s]{r'}{'},
    morestring=[s]{r"}{"},%
    morestring=[s]{r'''}{'''},%
    morestring=[s]{r"""}{"""},%
    morestring=[s]{u'}{'},
    morestring=[s]{u"}{"},%
    morestring=[s]{u'''}{'''},%
    morestring=[s]{u"""}{"""},%
    identifierstyle=\color{black}\ttfamily,
    commentstyle=\color{slha_cyan}\ttfamily,
    stringstyle=\color{slha_red}\ttfamily,
    keepspaces=true,
    showspaces=false,
    showstringspaces=false,
    rulecolor=\color{slha_frame},
    frame=single,
    frameround={t}{t}{t}{t},
    framexleftmargin=6mm,
    numbers=left,
    numberstyle=\tiny\color{halfgray},
    backgroundcolor=\color{slha_bg},
    basicstyle=\footnotesize,
    keywordstyle=\color{slha_green}\ttfamily,
    aboveskip=1.2em,
    belowskip=1.2em,
}

\lstdefinelanguage{MathIn}{
    morekeywords={Simplify,Eigenvalues},%
    emph={Start,InitUnitarity,GetScatteringDiagrams,BuildScatteringMatrix,MakeSPheno},%
    emphstyle={\color{mathematica_purple}},
    sensitive=true,%
    morecomment=[l]\%,%
    morestring=[b]',%
    morestring=[b]",%
    morestring=[s]{'''}{'''},
    morestring=[s]{"""}{"""},
    morestring=[s]{r'}{'},
    morestring=[s]{r"}{"},%
    morestring=[s]{r'''}{'''},%
    morestring=[s]{r"""}{"""},%
    morestring=[s]{u'}{'},
    morestring=[s]{u"}{"},%
    morestring=[s]{u'''}{'''},%
    morestring=[s]{u"""}{"""},%
    identifierstyle=\color{black}\ttfamily,
    commentstyle=\color{mathematica_cyan}\ttfamily,
    stringstyle=\color{mathematica_red}\ttfamily,
    keepspaces=true,
    showspaces=false,
    showstringspaces=false,
    rulecolor=\color{mathematica_frame},
    frame=single,
    frameround={t}{t}{t}{t},
    framexleftmargin=10mm,
    numbers=left,
    numberstyle=\tiny\color{halfgray},
    backgroundcolor=\color{mathematica_bg},
    basicstyle=\footnotesize,
    keywordstyle=\color{mathematica_green}\ttfamily,
    aboveskip=1.2em,
    belowskip=1.2em,
}

\lstdefinelanguage{MathOut}{
    morekeywords={Simplify,Eigenvalues},%
    sensitive=true,%
    morecomment=[l]\%,%
    morestring=[b]',%
    morestring=[b]",%
    morestring=[s]{'''}{'''},
    morestring=[s]{"""}{"""},
    morestring=[s]{r'}{'},
    morestring=[s]{r"}{"},%
    morestring=[s]{r'''}{'''},%
    morestring=[s]{r"""}{"""},%
    morestring=[s]{u'}{'},
    morestring=[s]{u"}{"},%
    morestring=[s]{u'''}{'''},%
    morestring=[s]{u"""}{"""},%
    identifierstyle=\color{black}\ttfamily,
    commentstyle=\color{mathematica_cyan}\ttfamily,
    stringstyle=\color{mathematica_red}\ttfamily,
    keepspaces=true,
    showspaces=false,
    showstringspaces=false,
    rulecolor=\color{mathematica_frame},
    frame=single,
    frameround={t}{t}{t}{t},
    framexleftmargin=10mm,
    numbers=left,
    numberstyle=\tiny\color{halfgray},
    backgroundcolor=\color{mathematica_bg},
    basicstyle=\footnotesize,
    keywordstyle=\color{mathematica_green}\ttfamily,
    aboveskip=1.2em,
    belowskip=1.2em,
}

\let\origthelstnumber\thelstnumber
\makeatletter
\newcommand*\Suppressnumber{%
  \lst@AddToHook{OnNewLine}{%
    \let\thelstnumber\relax%
     \advance\c@lstnumber-\@ne\relax%
    }%
}

\newcommand*\Reactivatenumber{%
  \lst@AddToHook{OnNewLine}{%
   \let\thelstnumber\origthelstnumber%
   \advance\c@lstnumber\@ne\relax}%
}

\usepackage{slashed}

\usepackage{booktabs}

\usepackage{tabularx, multirow}

\usepackage[width=0.9\textwidth,margin={30pt,30pt},font=small,labelfont=bf]{caption}
\numberwithin{equation}{section}

\usepackage{hyperref}

\usepackage{xcolor}
\hypersetup{
    colorlinks,
    linkcolor={red!0!black},
    citecolor={blue!50!black},
    urlcolor={blue!80!black}
}

\usepackage{graphicx}
\usepackage{cancel}
\usepackage{cite}
\usepackage{color}
\usepackage{bm}
\usepackage{multirow}
\usepackage{xspace}

\def\twomat[#1,#2][#3,#4]{\left( \begin{array}{cc} #1 & #2 \\ #3 & #4 \end{array} \right)}
\def\twoa[#1,#2][#3,#4]{\left( \begin{array}{cc} #1 & #2 \\ #3 & #4 \end{array} \right)}
\def\nn{\nonumber}

\def\thv[#1,#2,#3]{\left( \begin{array}{c} #1 \\ #2 \\ #3 \end{array} \right)}
\def\twv[#1,#2]{\left( \begin{array}{c} #1 \\ #2 \end{array} \right)}

\def\GeV{\ensuremath{\mathrm{GeV}}\xspace}
\def\TeV{\ensuremath{\mathrm{TeV}}\xspace}

\def\BSMArt{{\sc BSMArt}\xspace}
\def\SARAH{{\sc SARAH}\xspace}

\def\MicrOMEGAs{{\sc MicrOMEGAs}\xspace}
\def\SModelS{{\sc SModelS}\xspace}
\def\ZPEED{{\sc ZPEED}\xspace}
\def\ZPX{{\sc Z${}^\prime$-Explorer}\xspace}
\def\ov{\overline}
\def\bra{\langle}
\def\ket{\rangle}

\def\beq{\begin{equation}}
\def\eeq{\end{equation}}
\def\beqa{\begin{eqnarray}}
\def\eeqa{\end{eqnarray}}
\def\ba{\begin{eqnarray}}
          \def\ea{\end{eqnarray}}

\def\nn{\nonumber}

\title{ }
\author{Alon E Faraggi }
\date{August 2023}

\begin{document}

\begin{center}
{\Large \bf $M_W$ in String Derived Z$^\prime$ Models }
\end{center}
\vspace{2mm}
\begin{center}
{\large Alon E. Faraggi$^{(a)}$, Mark D. Goodsell$^{(b)}$ }\\
\vspace{2mm}

\it{$^{(a)}$Department of Mathematical Sciences, University of Liverpool, Liverpool L69 7ZL, UK}

\it{$^{(b)}$ Sorbonne Universit{\'e}, CNRS, Laboratoire de Physique Th{\'e}orique et Hautes Energies
(LPTHE), \\ F-75005 Paris, France}

\end{center}

\begin{abstract}
We introduce a phenomenological model for a string-derived $Z'$ scenario, and study its predictions for the mass of the W boson. In the process, we compare it to collider constraints for both pair-produced particles, Higgs boson properties, and $Z'$ searches. We also describe the implementation of new tools in the scanning code \BSMArt. 
\end{abstract}

\vfill\eject

\section{Introduction}

The Standard Model of particle physics provides a viable effective 
parameterisation of all observational data to date. Furthermore, 
the logarithmic evolution of the Standard Model parameters suggests that it
may provide viable perturbative parameterisation up to the Grand Unified 
Theory (GUT) scale or the Planck scale. It is nevertheless in general 
anticipated that the Standard Model is augmented by additional symmetries
and particles due to its many shortcomings, {\it e.g.}: it does not 
explain the origin of neutrino masses and the vast separation of 
fermion mass scales. But perhaps most importantly, the lightness 
of the Standard Model scalar sector compared to the GUT 
and Planck scales is not protected from radiative corrections
from the higher scales. Among the leading Beyond the Standard 
Model (BSM) theories that offers a remedy to this conundrum 
is supersymmetry. However, while supersymmetry can 
address the Standard Model hierarchy problem at the 
perturbative multi--loop level, it in fact reintroduces
it at leading order as a parameter in the superpotential. 
Supersymmetry mandates the existence of two doublet Higgs 
representation that in the simplest models
appear in vector--like representations and can receive a bi--linear 
mass term in the superpotential, the $\mu$--term. In the Minimal 
Supersymmetric Standard Model (MSSM) the $\mu$--parameter is set by 
hand to be of the order of the TeV scale, assuming in effect
the existence of some global or discrete symmetry that protects 
it from obtaining a larger value. However, from the point of view of 
fundamental theories that aim to unify the Standard Model with 
gravity such symmetries should emerge from the fundamental 
theory. Furthermore, it is well known that global symmetries 
are not preserved in theories of quantum gravity, and that 
is indeed the case in quasi--realistic string constructions. 
A possible solution to this problem is the existence 
of  an additional $U(1)$ symmetry that remains
unbroken down to the TeV scale and under which the 
Higgs doublets are chiral. However, the construction of 
string models that can accommodate such a $U(1)$ symmetry is hampered
by the fact that the extra $U1(1)$ symmetries that are typically
discussed in the Beyond the Standard Model (BSM) literature 
are anomalous in string derived models and cannot remain unbroken 
down to low scales. 
A string model that produces an anomaly free extra $U(1)$ symmetry 
and therefore can remain unbroken down to the TeV scale was constructed 
in \cite{Faraggi:2014ica}.

Anomaly cancellation down to the TeV scale in the $Z^\prime$ string derived model 
mandates the existence of additional states that are chiral under $U(1)_{Z^\prime}$. 
Some of these states are in charged vector--like representations under the Standard Model
gauge group factors, whereas other states are Standard Model singlets and can
serve as sterile neutrinos. 
Mass terms for these state can only be generated by the VEV that breaks the 
extra $U(1)$ symmetry. In turn the additional states can affect the experimentally 
measured values of the Standard Model parameters. In this paper our main focus 
is on the mass of the charged vector bosons $M_{W^\pm}$. This mass
parameter is related to the Standard Model $\rho$ parameter, which is 
measured experimentally to a high precision. It should be noted that at present there
is controversy with regard to the experimental measurement, where the CDF 
experiment at Fermilab reported a substantial deviation from the 
Standard Model value -- the combined Tevatron+LEP measurement is $80424 \pm 9$ MeV, compared to the SM value of $80356 \pm 6$ MeV -- whereas it disagrees by 4 standard deviations with (less precise, but in agreement with the SM) measurements by ATLAS and LHCb. We can therefore take a ``world average'' value of $80411 \pm 15$ MeV.

$M_W$ is particularly
interesting in the string derived $Z^\prime$ model because the
model predicts the existence of at least three additional pairs 
of Electroweak doublets whose mass is generated by the VEV that 
breaks the $Z^\prime$ symmetry. The masses of these extra doublets 
depend on the details of their couplings to the scalar field that
breaks the $Z^\prime$ symmetry. Given that some of these couplings 
can be small, the additional doublets may be relatively light compared
to the $Z^\prime$ breaking scale. Additionally, the existence of a 
multiplicity of states may enhance the effect of a single state. 
We shall determine precisely to what extent this is true in the following.

\section{The String Model~\label{string-model}}

In this section we elaborate on the construction of string 
derived models with additional $U(1)$ symmetries that may 
be detected in contemporary experiments. 
String inspired $Z^\prime$ phenomenology has been of interest since 
it was observed in the mid--eighties that string theory gives rise 
to models that reproduce the matter states and gauge interactions that 
appear in Grand Unified Theories (GUTs) like $E_6$ GUTs. However, the
construction of string derived models with such additional $U(1)$ symmetries
proved to be more intricate because they typically turn out to be anomalous
in the string models and hence cannot remain unbroken down to low scales. In string 
derived models the symmetry breaking pattern $E_6\rightarrow SO(10)\times U(1)_A$ results 
in non--vanishing trace for $U(1)_A$ due to the fact that some of the components in the 
chiral matter representations of $E_6$ are projected out from the massless spectrum, 
producing an anomalous $U(1)_A$. Attempts to construct string derived $Z^\prime$ models
that are not embedded in $E_6$ were discussed in the literature, but it is found that
agreement with the values of $\sin^2\theta_W(M_Z)$ and $\alpha_s(M_Z)$ favours the 
$E_6$ embedding in extra $U(1)$ models. We further note that the 
string derived $Z^\prime$ model possess an extended non--Abelian symmetry, beyond
the Standard Model at the string scale. The Higgs states available to break 
this extended non--Abelian symmetry is limited in the string models. This 
entails that the additional $U(1)$ symmetry, beyond the Standard Model, 
which is embedded in $SO(10)$, is broken in the string models at a scale that 
coincides with the non--Abelian breaking scale. Construction of string models with anomaly
free $U(1)_A$ can be obtained by following two routes. One is to extend $U(1)_A$ to a non--Abelian 
symmetry {\`a} la ref. \cite{Bernard:2012vf}, and the second is to utilise self--duality 
under the Spinor--Vector Duality (SVD), as was done in ref. \cite{Faraggi:2014ica}, which is the model
that we discuss here. In this model, the observable and hidden sector gauge symmetries are given by: 
\begin{eqnarray}
{\rm observable} ~: &~~SO(6)\times SO(4) \times
U(1)_1 \times U(1)_2\times U(1)_3 \nonumber\\
{\rm hidden}     ~: &SO(4)^2\times SO(8)~~~~~~~~~~~~~~~~~~~~~~~~~~~~~~~\nonumber
\end{eqnarray}
and the Higgs states available to break the non--Abelian gauge symmetry are 
\begin{align}
\overline{\cal H}({\bf\bar4},{\bf1},{\bf2})& \rightarrow u^c_H\left({\bf\bar3},
{\bf1},\frac 23\right)+d^c_H\left({\bf\bar 3},{\bf1},-\frac 13\right)+\nonumber\\
         &   ~~~~~~~~~~~~~~~~~~  {\overline {\cal N}}\left({\bf1},{\bf1},0\right)+
                             e^c_H\left({\bf1},{\bf1},-1\right)
                             \nonumber \\
{\cal H}\left({\bf4},{\bf1},{\bf2}\right) &
\rightarrow  u_H\left({\bf3},{\bf1},-\frac 23\right)+
d_H\left({\bf3},{\bf1},\frac 13\right)+\nonumber\\
         &  ~~~~~~~~~~~~~~~~~~   {\cal N}\left({\bf1},{\bf1},0\right)+
e_H\left({\bf1},{\bf1},1\right)\nonumber
\end{align}
The $U(1)$ symmetries $U(1)_1$ and $U(1)_2$ have a non--vanishing trace in the string 
model
\begin{equation}
{\rm Tr}U(1)_1= 36 ~~~~~~~{\rm and}~~~~~~~{\rm Tr}U(1)_3= -36.
\label{u1u3}
\end{equation}
and the combination which is embedded in $E_6$, 
\begin{equation}
U(1)_\zeta ~=~ U(1)_1+U(1)_2+U(1)_3~,
\label{uzeta}
\end{equation}
is anomaly free
and can be part of an unbroken $U(1)_{Z^\prime}$ symmetry down to low scales.
The Vacuum Expectation Values of ${\cal N}$ and $\overline{\cal N}$ along flat $F$-- and 
$D$--directions leaves $N=1$ spacetime supersymmetry and the $U(1)$ combination
\begin{equation}
U(1)_{{Z}^\prime} ~=~
\frac{1}{5} (U(1)_C - U(1)_L) - U(1)_\zeta
~\notin~ SO(10),
\label{uzpwuzeta}
\end{equation}
unbroken below the string scale. 
The $U(1)$ symmetry in eq. (\ref{uzpwuzeta}) is anomaly free 
if $U(1)_\zeta$ is anomaly free. Anomaly cancellation down to low scale 
requires the existence of additional chiral states in the spectrum that are 
vector--like with respect to the Standard Model gauge group. The spectrum below the 
non--Abelian symmetry breaking scale is shown in table \ref{table27rot}. 

\begin{table}[!ht]
\noindent
{\small
\begin{center}
{
\begin{tabular}{|l|cc|c|c|c|}
\hline
Field &$\hphantom{\times}SU(3)_C$&$\times SU(2)_L $
&${U(1)}_{Y}$&${U(1)}_{Z^\prime}$  \\
\hline
$\hat{Q}_L^i$&    $3$       &  $2$ &  $+\frac{1}{6}$   & $-\frac{2}{5}$   ~~  \\
$\hat{u}_L^i$&    ${\bar3}$ &  $1$ &  $-\frac{2}{3}$   & $-\frac{2}{5}$   ~~  \\
$\hat{d}_L^i$&    ${\bar3}$ &  $1$ &  $+\frac{1}{3}$   & $-\frac{4}{5}$  ~~  \\
$\hat{e}_L^i$&    $1$       &  $1$ &  $+1          $   & $-\frac{2}{5}$  ~~  \\
$\hat{L}_L^i$&    $1$       &  $2$ &  $-\frac{1}{2}$   & $-\frac{4}{5}$  ~~  \\
\hline
$\hat{D}^i$       & $3$     & $1$ & $-\frac{1}{3}$     & $+\frac{4}{5}$  ~~    \\
$\hat{{\bar D}}^i$& ${\bar3}$ & $1$ &  $+\frac{1}{3}$  &   $+\frac{6}{5}$  ~~    \\
$\hat{H}_{\textrm{vl}}^i$       & $1$       & $2$ &  $-\frac{1}{2}$   &  $+\frac{6}{5}$ ~~    \\
$\hat{{\bar H}}_{\textrm{vl}}^i$& $1$       & $2$ &  $+\frac{1}{2}$   &   $+\frac{4}{5}$   ~~  \\
\hline
$\hat{S}^i$       & $1$       & $1$ &  ~~$0$  &  $-2$       ~~   \\
\hline
$\hat{H}_1$         & $1$       & $2$ &  $-\frac{1}{2}$  &  $-\frac{4}{5}$  ~~    \\
$\hat{H}_2$  & $1$       & $2$ &  $+\frac{1}{2}$  &  $+\frac{4}{5}$  ~~    \\
\hline
$\hat{\phi}$       & $1$       & $1$ &  ~~$0$         & $-1$     ~~   \\
$\hat{\bar\phi}$       & $1$       & $1$ &  ~~$0$     & $+1$     ~~   \\
\hline
\hline
$\hat{\zeta}^i$       & $1$       & $1$ &  ~~$0$  &  ~~$0$       ~~   \\
\hline
\end{tabular}}
\end{center}
}
\caption{\label{table27rot}
Supermultiplet spectrum and
$SU(3)_C\times SU(2)_L\times U(1)_{Y}\times U(1)_{{Z}^\prime}$
quantum numbers, with $i=1,2,3$ for the three light
generations. The charges are displayed in the
normalisation used in free fermionic
heterotic--string models. }
\end{table}

\section{Phenomenological Models}
\label{SEC:Models}

Having defined the field content of the model, in principle we can write down the Lagrangian. From this, the pattern of symmetry breaking can be determined and the mixing of the different fields determined. However, it is possible to impose on the theory further (global) symmetries, potentially only approximate, that will reduce the number of possible couplings and split the final sets of fields into smaller multiplets. This should be regarded as additions to the R-parity of the MSSM which prevents, inter alia, the $H_d$ fields mixing with the lepton multiplets. 

With so many additional fields there are a plenitude of possible choices. Models are possible where the fields $H_{2I}, H_{1I}$ are Higgs-like and, by giving their neutral components expectation values, we could have one Higgs per generation. For the time being this is technically challenging to explore, and we postpone it to future work. 

An alternative is to assign a lepton number to these fields; as in \cite{Faraggi:2022emm} lepton number should be explicitly broken by a neutrino mass, but for the purposes of computing collider constraints this could be neglected. We describe such a model 
in appendix \ref{APP:LeptonNo}. However, the constraints on new leptons are now rather severe, and we shall not focus on that model in this work as the bounds push new states to be sufficiently heavy that they can have little impact on the W boson mass. 

In \cite{King:2007uj,Athron:2011ew} a version of the $E_6$ theory with a discrete $Z_{2H}$ was presented. It has the advantage of simplifying the model and restricting the number of interactions. All of the new fields beyond the MSSM are charged under the new symmetry, with the exception of one singlet, which obtains an expectation value to break the $U(1)_{Z'}$. This leads to two sectors of the theory with separate lightest particles that cannot decay from one to the other. There are also two spectator singlet fields that have the same charges under this symmetry and we ought therefore to split their neutral scalar components into real and imaginary parts after EWSB because nothing keeps them as complex fields. This is somewhat cumbersome to implement in a numerical code and we shall not investigate it here: instead below we present a variation based on a continuous symmetry (which should be regarded as approximate, similar to lepton number in the SM).

\subsection{$U(1)_H$ Model}

By assigning a continuous symmetry to the additional fields we ensure that the neutral components of the fields remain complex scalars even after electroweak symmetry breaking. This mere technicality is of practical importance given that we have three generations of additional doublets, and enables the simplification of the model. 

\begin{table}[h]
\begin{center}
{\small 
Chiral and gauge multiplet fields  of the MSSM\\
\begin{tabular}{|c|c|c|c|c|c|c|}
\hline
  Superfield     & Generations   & Scalars                  & Fermions & Vectors & ($SU(3)$, $SU(2)$, $U(1)_Y$)  & $U(1)_{Z'}$   \\ \hline
$\mathbf{Q}_i$   & 3 & $\tilde{q}_i=(\tilde{u}_{i,L},\tilde{d}_{i,L})$  & $(u_L,d_L)$ & & (\textbf{3}, \textbf{2}, 1/6) & $-2/5$ \\ 
$\mathbf{U}_i$   & 3 & $\tilde{u}_{i,R}$              & $u_{i,R}$     & & ($\overline{\textbf{3}}$, \textbf{1}, -2/3) & $-2/5$ \\
$\mathbf{D}_i$   & 3 &$\tilde{d}_{i,R}$     & $d_{i,R}$     & & ($\overline{\textbf{3}}$, \textbf{1}, 1/3) & $-4/5$ \\ 
$\mathbf{L}_i$    & 3 &($\tilde{\nu}_{i,L}$,$\tilde{e}_{i,L}$) & $(\nu_{i,L},e_{i,L})$ & & (\textbf{1}, \textbf{2}, -1/2) & $-4/5 $ \\
$\mathbf{E}_i$   &3 & $\tilde{e}_{i,R}$    & $e_{i,R}$          & & (\textbf{1}, \textbf{1}, 1) & $-2/5$  \\ \hline
$\mathbf{H_u}$  & 1 &$(H_u^+ , H_u^0)$ & $(\tilde{H}_u^+ , \tilde{H}_u^0)$ & & (\textbf{1}, \textbf{2}, 1/2) & $4/5$  \\ 
$\mathbf{H_d}$  & 1 &$(H_d^0 , H_d^-)$ & $(\tilde{H}_d^0 , \tilde{H}_d^-)$ & & (\textbf{1}, \textbf{2}, -1/2) & $6/5$ \\ \hline
$\mathbf{W_{3,\alpha}}$ & 1 & & $\lambda_{3} $                       & $G_\mu$              & (\textbf{8}, \textbf{1}, 0) & 0 \\
$\mathbf{W_{2,\alpha}}$ & 1 & & $\tilde{W}^0, \tilde{W}^\pm $ & $W^{\pm}_\mu , W^0_\mu$  & (\textbf{1}, \textbf{3}, 0) & 0 \\ 
$\mathbf{W_{Y,\alpha}}$ & 1 & & $ \tilde{B} $                       & $B_\mu$              & (\textbf{1}, \textbf{1}, 0 ) & 0  \\ 
\hline
\end{tabular}}\\[4mm]
{\small Additional chiral and gauge multiplet fields in the {\tt E6AFH} model}\\
\begin{tabular}{|c|c|c|c|c|c|c|c|}
\hline
  Superfield               & Generations  &Scalars                 & Fermions & Vectors &  & $U(1)_{Z'}$ & $U(1)_H$\\ \hline
$\mathbf{W_{Z',\alpha}}$ & 1& & $\tilde{B}'$ & $B^\prime_\mu$ & (\textbf{1}, \textbf{1}, 0 ) & 0 & 0 \\ \hline
  $\mathbf{S}_0$&1& $S_0$ & $F_{S}$   & & (\textbf{1},\textbf{1},0) & -2 & 0 \\
  $\mathbf{S}_P$&1& $S_P$ & $F_{SP}$   & & (\textbf{1},\textbf{1},0) & -2 & 1\\
  $\mathbf{S}_M$&1& $S_M$ & $F_{SM}$   & & (\textbf{1},\textbf{1},0) & -2 & -1\\
    $\mathbf{\phi}$&1& $ \Phi$ & $F_{\phi}$   & & (\textbf{1},\textbf{1},0) & 1 & 1 \\
$\mathbf{\tilde{\phi}} $&1& $ \tilde{\Phi}$ & $F_{\tilde{\phi}}$   & & (\textbf{1},\textbf{1},0) & -1 & -1 \\
  $\mathbf{X_d}$&1 & $(X_d^0 , X_d^-)$ & $(\tilde{X}_d^0 , \tilde{X}_d^-)$ & & (\textbf{1}, \textbf{2}, -1/2) & $-4/5$ & 0\\
  $\mathbf{X_u}$&1  & $(X_u^+ , X_u^0)$ & $(\tilde{X}_u^+ , \tilde{X}_u^0)$ & & (\textbf{1}, \textbf{2}, 1/2) & $4/5$ & 0\\
  $\mathbf{H_{1 I=\{1,2\}}}$ &2 & $(H_{1I}^0 , H_{1I}^-)$ & $(\tilde{H}_{1I}^0 , \tilde{H}_{1I}^-)$ & & (\textbf{1}, \textbf{2}, -1/2) & $6/5$ & 1\\
  $\mathbf{H_{2 I=\{1,2\}}}$ &2 & $(H_{2I}^+ , H_{2I}^0)$ & $(\tilde{H}_{2I}^+ , \tilde{H}_{2I}^0)$ & & (\textbf{1}, \textbf{2}, 1/2) & $4/5$ & -1 \\
  $\mathbf{D_x}$&1 & $ D_x$ & $F_{D_x} $ & & (\textbf{3}, \textbf{1}, -1/3) & $4/5$ & 0 \\
   $\mathbf{\tilde{D}_x}$&1 & $ \tilde{D}_x$ & $F_{\tilde{D}_x} $ & & ($\mathbf{\ov{3}}$, \textbf{1}, 1/3) & $6/5$ & 0\\
\hline
\end{tabular}
\caption{\label{TAB:U1fields}Field content of the lepton model, as implemented in \SARAH as {\tt E6AFH}.  
Top panel: chiral and gauge multiplet fields of the MSSM with their $U(1)$ charges; bottom panel: additional fields and their $U(1)_H$ charges. All MSSM fields are neutral under $U(1)_H$.}
\end{center}
\end{table}


As in \cite{King:2007uj,Athron:2011ew}, we assign lepton number to the additional vector-like doublet, so that they mix with the electrons and give a new pair of heavy neutrinos; unlike those references we do not include additional right-handed neutrino mulitplets.  The subequent field content of the theory is given in table \ref{TAB:U1fields}. The most general superpotential consistent with these symmetries is rather simple:
\begin{align}
  W =& Y_u u q H_u - Y_d d q H_d - Y_e e l  H_d  + \mu X_u X_d + \lambda S_0 H_u H_d + \lambda_{12}^{IJ} S_0 H_{2I} \cdot H_{1J} \nn\\
     &+ \kappa S_0 D_x D_{\ov{x}} + g_{QL} D_x q q + g_{QR} D_{\ov{x}} d u + \mu^\prime \phi \tilde{\phi}\nn\\
  &+ \lambda_5 S_M H_u H_{1I} + \lambda_6 S_P H_{2I} H_d +Y_x e X_d H_d.
\end{align}
After EWSB, the neutral and charged components of the fields are no longer degenerate, and there are many mixings among them. The resulting sets of fields and their composition in terms of pre-EWSB eigenstates is given in table \ref{TAB:MixingsU1}.
\begin{table}[h] \centering
\begin{tabular}{|c|c|c|c|}
\hline
  Multiplet Description   & Symbol & Number  of fields  & Gauge eigenstates    \\ \hline
  Neutralinos & $\tilde{\chi}^0$& 6 & $ \tilde{B}, \tilde{W}^0, \tilde{H_d^0}, \tilde{H_u^0}, F_{S}, \tilde{B}^\prime$ \\
  Charginos & $\tilde{\chi}^-$ &2 & $(\tilde{W}^+, \tilde{h}_d^0),(\tilde{W}^-, \tilde{h}_u^0 )$ \\
  Charged leptons & $e^-$ & 4 & $(e_{i,L}, \tilde{X}_d^-), (e_{i,R}, \tilde{X}_u^+)$ \\
  Heavy (dirac) neutrinos & $F_{DX}$& 1 & $(\tilde{X}_d^0), (\tilde{X}_u^0)$ \\
  Selectrons & $\tilde{e}$ & 8 & $\tilde{e}_{i,L}, (\tilde{e}_{i,R})^*, X_d^-, (X_d^+)^*$ \\
  Sterile charged scalars  & $S_{HI}^-$& 2 &$H_{1I}^-, (H_{2I}^+)^*$ \\
  Sterile neutral scalars & $S_{HI}^0 $ & 6 & $H_{1I}^0, (H_{2I}^0)^*, S_P, (S_M)^*,\Phi, \tilde{\Phi}^*$\\
  Sterile (dirac) neutralinos & $\tilde{\chi}^{\prime 0}$& 4 & $(\tilde{H}_{1I}^0,F_{S_P}, F_{\phi}), (\tilde{H}_{2I}^0, F_{S_M}, F_{\tilde{\phi}})$ \\
  Sterile charginos & $\tilde{\chi}^{\prime +}$& 2 & $(\tilde{H}_{1I}^-), \tilde{H}_{2I}^+)$ \\
  Sneutrinos & $\tilde{\nu}$ & 5 & $\tilde{\nu}_L, X_d^0, (X_u^0)^*$ \\
\hline
\end{tabular}
\caption{\label{TAB:MixingsU1} Mixing of gauge eigenstate fields into mass eigenstates for the $U(1)_H$ model.}
\end{table}

There are several salient phenomenological features. Firstly, the $U(1)_H$ symmetry splits the field content into two sectors, which do not mix; nor can the sector charged under the new symmetry decay exclusively to the neutral sector; thus, there is an LSP in the charged sector. This is familiar to the case of the $Z_H$ symmetry in \cite{King:2007uj,Athron:2011ew}. 

As a consequence, the mass matrices of the sector neutral under the  $U(1)_H$ (with the exception of the leptons, which have an additional vector-like pair) are the same as in the minimal $U(1)$ extension of the MSSM \cite{Fayet:1974pd,Kim:1983dt,Cvetic:1995rj,Cvetic:1997ky}. Constraints from direct searches for $Z'$ bosons push its mass to be in the multi-TeV range (we shall place precise limits in our model below) because it has a sizeable branching fraction to leptons; since the mass of the $Z'$ is roughly $2 v_S g_X$, where $\bra S_0 \ket \equiv v_S/\sqrt{2}$, $g_X$ is the new gauge coupling (the factor $2$ coming from the singlet's charge), then the singlet expectation value must be substantial. 

The properties of the new `sterile' scalars and fermions will therefore be of most interest in this work: we shall examine constraints arising from them and to what extent they can be compatible with an enhanced mass for the $W$ boson. Examining first the sterile scalars, if we make the approximation that their mass terms and Yukawa couplings $m_{H_{1I}}^2,m_{H_{2I}}^2, \lambda_{12}$ are diagonal and neglect SUSY-breaking trilinear couplings and kinetic mixing, and ignoring the $\phi, \tilde{\phi}$ states because they do not mix or have other interactions, then for the neutral scalars we have
\begin{align}
&\mathcal{M}_{S_{HI}^0}^2 \sim \\
&\left( \begin{array}{cccc} 
m_{S_P}^2 + 2 g_X^2 v_S^2 & 0 & \mathcal{O} (v^2) & \mathcal{O} (v^2) \\ 
0 & m_{S_M}^2 + 2 g_X^2 v_S^2 & \mathcal{O} (v^2) & \mathcal{O} (v^2) \\
\mathcal{O} (v^2) & \mathcal{O} (v^2) & m_{S_{HI}\,11}^2 - \frac{6}{5} g_X^2 v_S^2 + \frac{1}{2} v_S^2 (\lambda_{12}^{11})^2 + \mathcal{O} (v^2)& \mathcal{O} (v^2) \\
\mathcal{O} (v^2) & \mathcal{O} (v^2) & \mathcal{O} (v^2)& m_{S_{HI}\,22}^2 - \frac{6}{5} g_X^2 v_S^2 + \frac{1}{2} v_S^2 (\lambda_{12}^{22})^2 + \mathcal{O} (v^2) 
\end{array} \right) \nn
\end{align}
while for the charged scalars we have two copies of 
\begin{align}
\mathcal{M}_{S_{HI}^-}^2 \sim& \left( \begin{array}{cc} 
m_{S_{HI}\,ii}^2 - \frac{6}{5} g_X^2 v_S^2 + \frac{1}{2} v_S^2 (\lambda_{12}^{ii})^2 + \mathcal{O} (v^2) & \mathcal{O} (v^2) \\
\mathcal{O} (v^2) & m_{S_{HI}\,ii}^2 - \frac{4}{5} g_X^2 v_S^2 + \frac{1}{2} v_S^2 (\lambda_{12}^{ii})^2 + \mathcal{O} (v^2) 
\end{array} \right) .
\end{align}
If we would like to use new states to enhance the $W$-boson mass at one loop, then we need them to be light and to have their masses substantially split by electroweak symmetry breaking. If we take a $Z'$ boson at $5$ TeV, however, then we see that if we have one lighest charged scalar at $\mathcal{O}(100)$ GeV, the heavier charged scalar will be at roughly $1500$ GeV, and so the mixing among the charged scalars will only be small (since the mixing between $H_{1I}^-$ and $H_{2I}^-$ is proportional to the off-diagonal couplings). Simliarly in the neutral sector, the coupling to $v$ is via $\lambda_5, \lambda_6$ and through mixing with the $S_P, S_M$ states. Hence the only way to obtain sizeable mixings would be to arrange for tachyonic masses for $m_{S_P}^2$ and/or $m_{S_M}^2$ such that two or more neutral scalars are near the electroweak scale. Since this requires some rather delicate fine-tuning we shall not explore such a possibility here, but leave it to future work.

For the sterlile charginos we have a mass matrix 
\begin{align}
\mathcal{M}_{\tilde{\chi}^\prime\, ij} =& \frac{v_S}{\sqrt{2}} \lambda_{12}^{ij}
\end{align}
whereas for the sterile (dirac) neutralinos we have
\begin{align}
    \mathcal{M}_{\tilde{\chi}^{\prime 0}} = \left( \begin{array}{cccc}
    - \frac{v_S}{\sqrt{2}} \lambda_{12}^{11} & - \frac{v_S}{\sqrt{2}} \lambda_{12}^{12} & - \frac{v}{\sqrt{2}} \cos \beta \lambda_6^1 & 0 \\
    - \frac{v_S}{\sqrt{2}} \lambda_{12}^{21} & - \frac{v_S}{\sqrt{2}} \lambda_{12}^{22} & - \frac{v}{\sqrt{2}} \cos \beta \lambda_6^2 & 0 \\
     - \frac{v}{\sqrt{2}} \sin \beta \lambda_5^1  & - \frac{v}{\sqrt{2}} \sin \beta \lambda_5^2 & 0 & 0 \\
     0 & 0 & 0& \mu^\prime
    \end{array} \right).
\end{align}
While the $\phi, \Phi$ states decouple in this approximation (until we break the $U(1)_H$ explicitly) the sterile neutralinos have a see-saw mechanism; for the case again of diagonal couplings $\lambda_{12}^{11} = \lambda_{12}^{22} \equiv \lambda_{12}, \lambda_5^1 = \lambda_5^2 \equiv \lambda_5, \lambda_6^1 = \lambda_6^2 \equiv \lambda_6,$ we have one state with mass $\frac{v_S}{\sqrt{2}} \lambda_{12}, $  and in the limit that $\lambda_{12} v_S \gg \lambda_5 v, \lambda_6 v$ the lightest state has mass $\frac{\sqrt{2} \lambda_5 \lambda_6 v^2 \sin \beta \cos \beta}{\lambda_{12} v_S},$ which can be very small and is certainly below the electroweak scale. In the opposite limit, the masses are bounded by $v \lambda_6 \cos \beta/\sqrt{2}, v \lambda_5 \sin \beta/\sqrt{2}$, but we cannot make $\lambda_{12}$ arbitrarily small because they control the sterile chargino masses, which must be above the LEP bound; and because there is always one neutral fermion with that mass. Moreover, we note that the lightest state is always suppressed by $\tan \beta.$

Hence, this model has a very light neutral fermion, and we must ensure that the Higgs boson does not substantially decay to it. The coupling of the Higgs to the fermions proceeds through the couplings $\lambda_5, \lambda_6,$ and hence the lightest sterile neutralino must either be somewhat heavier than half the Higgs mass, requiring small $\tan\beta$ and large couplings; or it will be rather light indeed. The latter case is therefore most typical, and the model will therefore lead to collider signatures with large amounts of missing energy from the neutral fermion decays. However, these fermions provide a good opportunity for enhancing the $W$ boson mass. 



\section{Constraints and numerical setup}

We shall consider constraints from:
\begin{itemize}
\item The Higgs boson mass and measured values of its couplings. 
\item LHC searches for heavy Higgs bosons.
\item LHC searches for pair-produced particles.
\item LHC $Z'$ searches.
\item Limits on $\mu \rightarrow 3e$ and $\mu \rightarrow e \gamma$.
\item Measurements of the $W$ boson mass. 
\end{itemize}

Direct searches for vector-like leptons now give stringent constraints; notably \cite{CMS:2019hsm} constrains the mass of a degenerate doublet to be above 790 GeV if the decay is $100\%$ to third-generation leptons. We could weaken this bound by considering the decays to the first two generations to be dominant, but the search would still provide some bound -- it is just not clear what that would be, since no recast is available. On the other hand, in \cite{Kawamura:2023zuo} two ATLAS searches were recast and robust bounds placed on such scenarios; the limits should be of the order of a $600$ GeV minimum mass for these vector-like leptons. Hence it is implausible that they could drive an enhancement of the W mass. We shall therefore consider our vector-like fermions to be heavy. 


To explore the parameter space of our model, we implemented our $U(1)_H$ model into \SARAH \cite{Staub:2013tta,Goodsell:2017pdq,Braathen:2017izn,Goodsell:2018tti,Goodsell:2020rfu,Benakli:2022gjn} (we also implemented the lepton-number version, as described in the appendix, but shall not consider it further). 
This produces a fortran code linked to the {\sc SPheno} \cite{Porod:2011nf} library, which can compute the spectrum, decays, and flavour processes including rare lepton number violating ones. 
We use \BSMArt \cite{Goodsell:2023iac} version 1.2, which is released to accompany this paper, and we give details of the new implementations in appendix \ref{APP:BSMArt}. In particular, the code has been expanded to compute LHC limits using \SModelS \cite{Kraml:2013mwa,Alguero:2021dig,MahdiAltakach:2023bdn}, for which cross-sections are computed using \MicrOMEGAs \cite{Belanger:2007zz,Belanger:2010pz,Belanger:2018ccd}; plus $Z'$ limits using \ZPEED\cite{Kahlhoefer:2019vhz} and \ZPX\cite{Alvarez:2020yim,Lozano:2021zbu}.

To look for viable models, we first set up a Markov-Chain Monte-Carlo (MCMC) scan in \BSMArt over GUT-scale parameters. We then set up refined scans of low-energy parameters around chosen benchmark points. We shall describe the parameter choices and the likelihood functions in the next section, along with the results.


\section{Results}

\subsection{High-scale boundary conditions}

To define a model at the GUT scale, we define a common gaugino mass $m_{1/2}$, common trilinear term $A_0$ and common scalar mass $m_0,$ as in the CMSSM. However, for certain scalars beyond the MSSM, it is necessary to have a somewhat larger supersymmetry-breaking mass-squared terms to avoid tachyons at low energies: we can see from the above mass-matrices that the expectation value of the singlet that breaks the $U(1)'$ symmetry drives the $H_{1I}, H_{2I}$ fields tachyonic, and in our model we want to avoid this. Hence we put  
\begin{align}
m_{H_{1I}}^2 = m_{H_{2I}}^2 = m_{S_P}^2 = m_{S_M}^2 = 4 m_0^2 \ \mathrm{at}\ m_{GUT}.
\end{align}
The factor $4$ is a convenient choice; in the input files for the \SARAH model we allow it to be a free parameter.

In addition, we take diagonal couplings:
\begin{align}
\lambda_{12}^{ij}\equiv& \lambda^\prime \delta^{ij} \nn\\
\lambda_5^i\equiv& \left(\begin{array}{c}\lambda_5 \\ \lambda_5 \end{array} \right)^i \nn\\
\lambda_6^i\equiv&\left(\begin{array}{c}\lambda_6 \\ \lambda_6 \end{array} \right)^i \nn\\
Y_X^i \equiv& (10^{-4}, 10^{-4}, 10^{-4})^i \nn\\
\kappa^{ij} \equiv& 0.1 \delta^{ij},
\end{align}
and we take $g_{QL}^{333} = g_{QR}^{333} = 0.1$ and zero otherwise; this latter is not important for this study, and the choice of $Y_X$ to be small ensures that the rare lepton-number violation processes are negligible. At the GUT scale we also fix $\mu = \mu^\prime = 3\TeV,$ since these terms are not phenomenologically relevant for this study. The only other input parameters -- which are defined at low energies -- are $v_S$, $\tan \beta.$ and $B_\mu \equiv b_0^2$ where $B_\mu$ is the soft term associated with the MSSM Higgs $\mathcal{L} \supset B_\mu H_u \cdot H_d$. Hence in total we have $10$ free parameters:
$$m_0, m_{12}, a_0, \tan \beta, v_S, \lambda, b_0, \lambda^\prime, \lambda_5, \lambda_6.$$
We then scan these within the ranges:
\begin{gather}
m_0 \in [2500,4000]\ \GeV, \ m_{1/2} \in [1500,3500]\ \GeV, \ A_0 \in [-3500,-200]\ \GeV,\nn\\
v_S \in [13000,30000]\ \GeV,\ b_0 \in [1100,2000]\ \GeV, \nn\\
\tan \beta \in [2,40],\ \lambda \in [0.01,0.1],\ \lambda^\prime \in [0.005,0.1],\ \lambda_5 \in [0.4,0.8],\ \lambda_6 \in [-0.008,0.008].
\end{gather}
To this end we implement a Markov-Chain Monte-Carlo (MCMC) scan in {\sc BSMArt}\cite{Goodsell:2023iac} using a likelihood function based on: the Higgs mass; the $W$ boson mass (a gaussian likelihood peaked at $80.410$ GeV with variance $30$ MeV); the likelihood from {\sc ZPEED} (we take the exponential of the log likelihood from that code), Higgs data from {\sc HiggsTools} \cite{Bahl:2022igd}: the $p$--value for the SM-like Higgs and, for the other (heavy) scalars and pseudoscalars, we construct a fake likelihood from the ratio of observed cross-section to the best limit; and we use the same fake likelihood for the ratio of observed cross-section from {\sc SModelS} {\tt v2.3.0}. Hence the scan samples predominantly around point of high likelihood which should favour those with a high $W$--boson mass (as close as possible to the world average post-CDF) and passing all constraints. 

\begin{figure}[h!]
    \centering
   \begin{subfigure}[b]{0.45\textwidth}
        \centering
        \includegraphics[width=\linewidth]{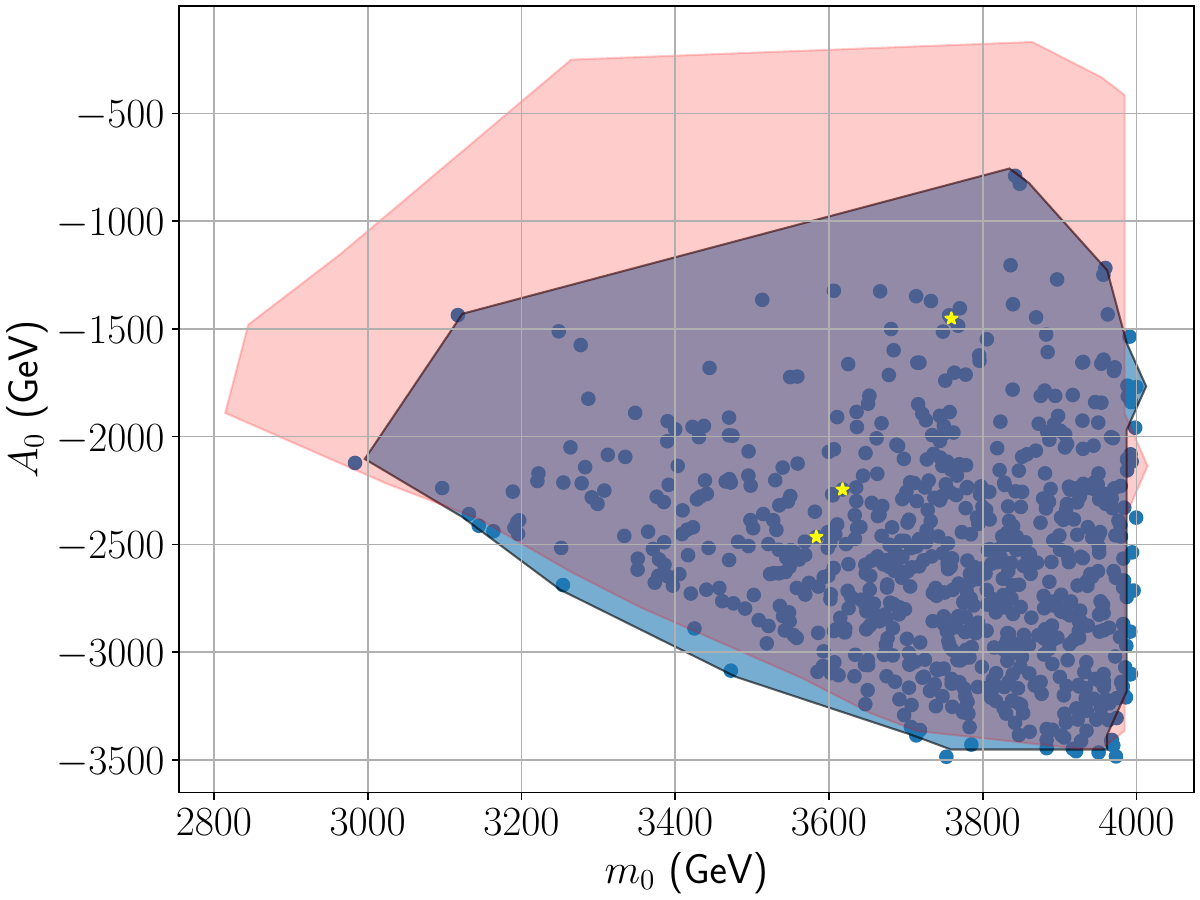} %
        \caption{}
    \end{subfigure}%
   \begin{subfigure}[b]{0.45\textwidth}
        \centering
        \includegraphics[width=\linewidth]{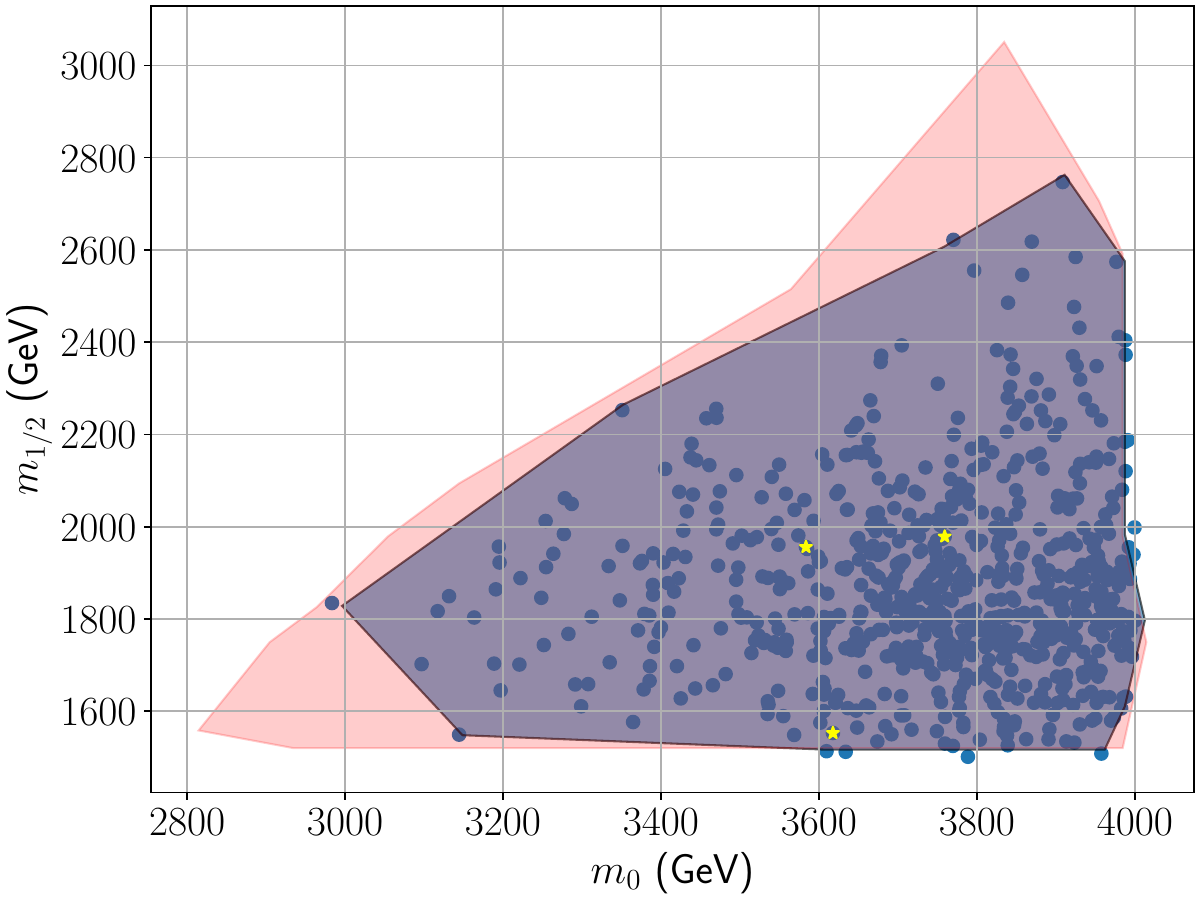}
        \caption{}
    \end{subfigure}
    \caption{\label{FIG:Highm0}Allowed parameter space for high-scale boundary conditions in terms of high-energy parameters $m_0, m_{1/2}, A_0$. Regions where unacceptable points were found are shown in red, acceptable points are shown in the blue region as blue dots. Our three best points are shown as yellow stars.}
\end{figure}

\begin{figure}[h!]
    \centering
   \begin{subfigure}[b]{0.45\textwidth}
        \centering
        \includegraphics[width=\linewidth]{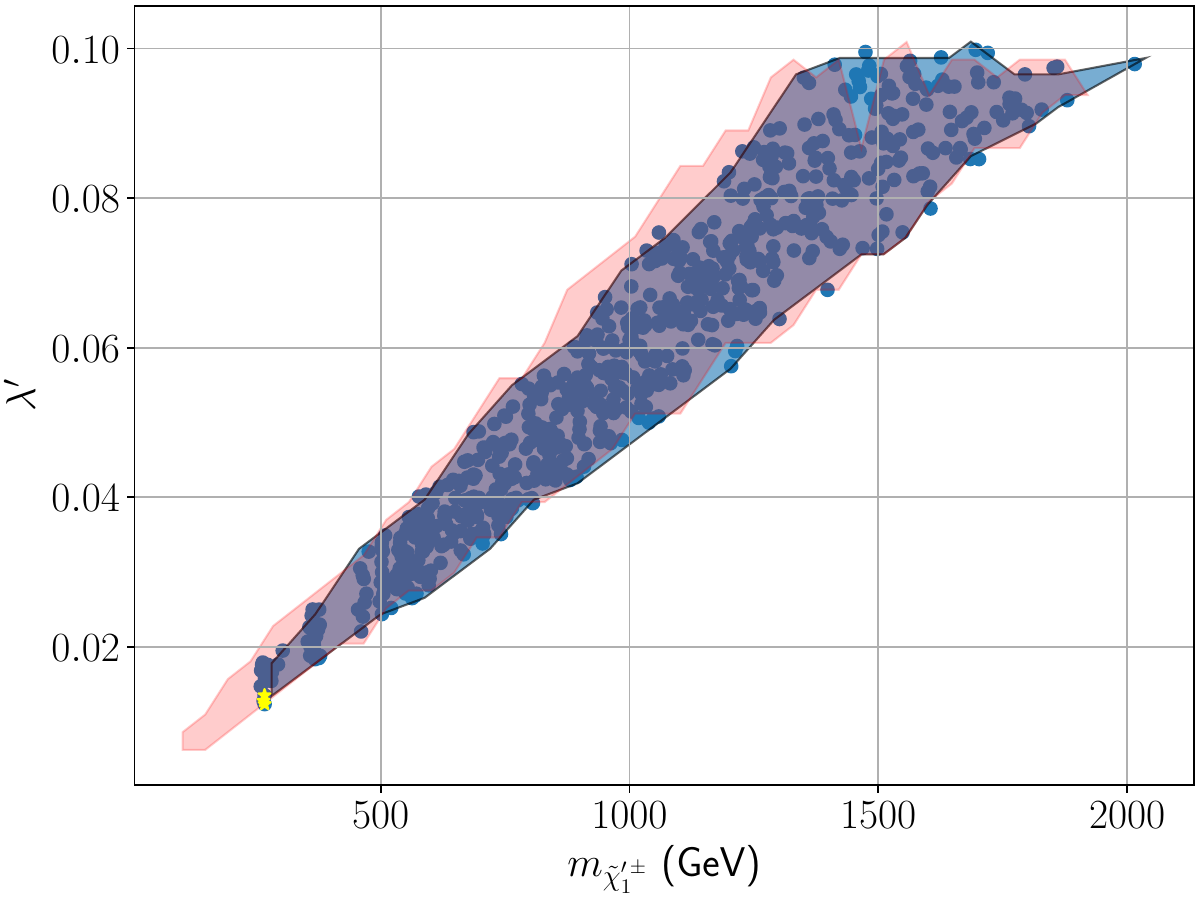} %
        \caption{}
    \end{subfigure}%
   \begin{subfigure}[b]{0.45\textwidth}
        \centering
        \includegraphics[width=\linewidth]{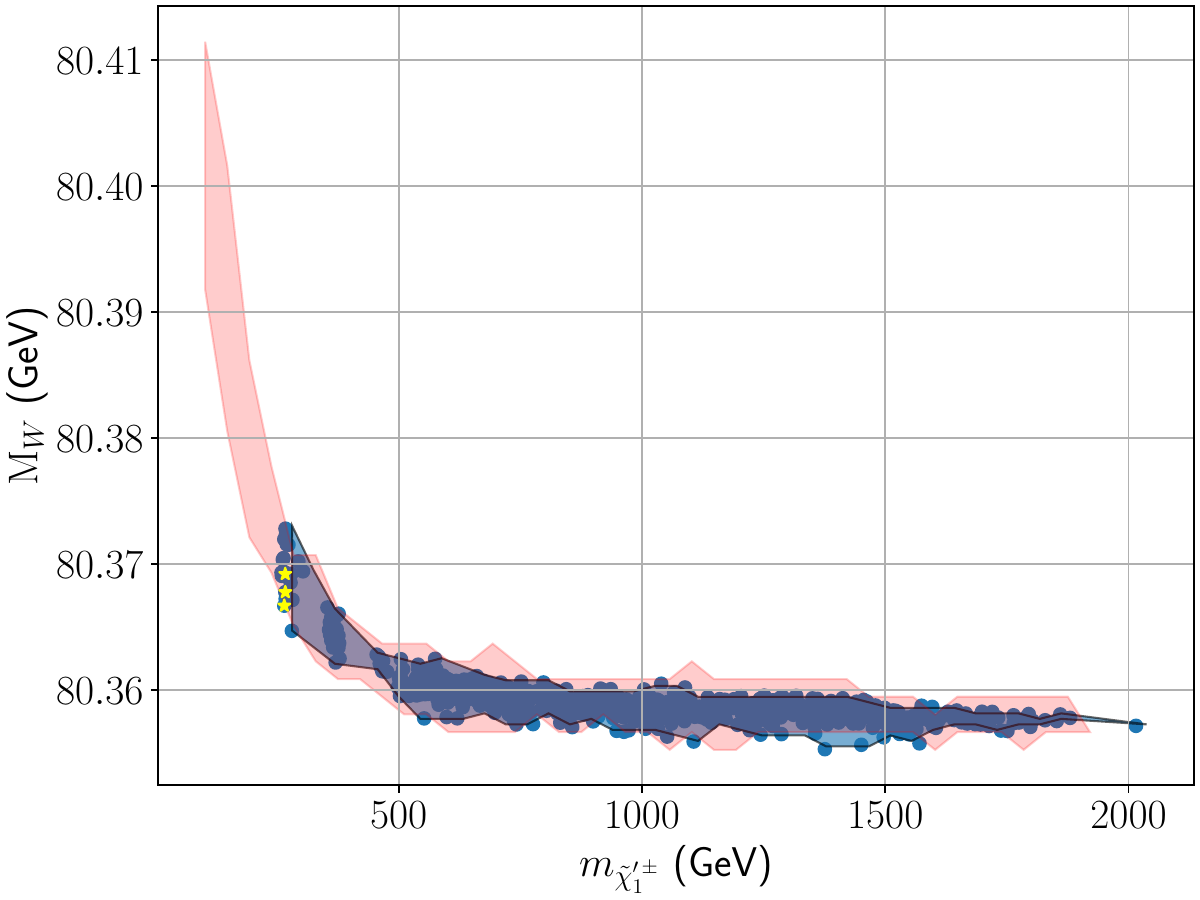}
        \caption{}
    \end{subfigure}
    \caption{\label{FIG:HighmChaP}Allowed parameter space for high-scale boundary conditions in terms of observables. (a) shows the proportionality of the sterile gaugino mass to $\lambda^\prime;$ (b) shows the dependence of the W boson mass on the sterile gaugino mass. }
\end{figure}

\begin{figure}[h!]
    \centering
   \begin{subfigure}[b]{0.45\textwidth}
        \centering
        \includegraphics[width=\linewidth]{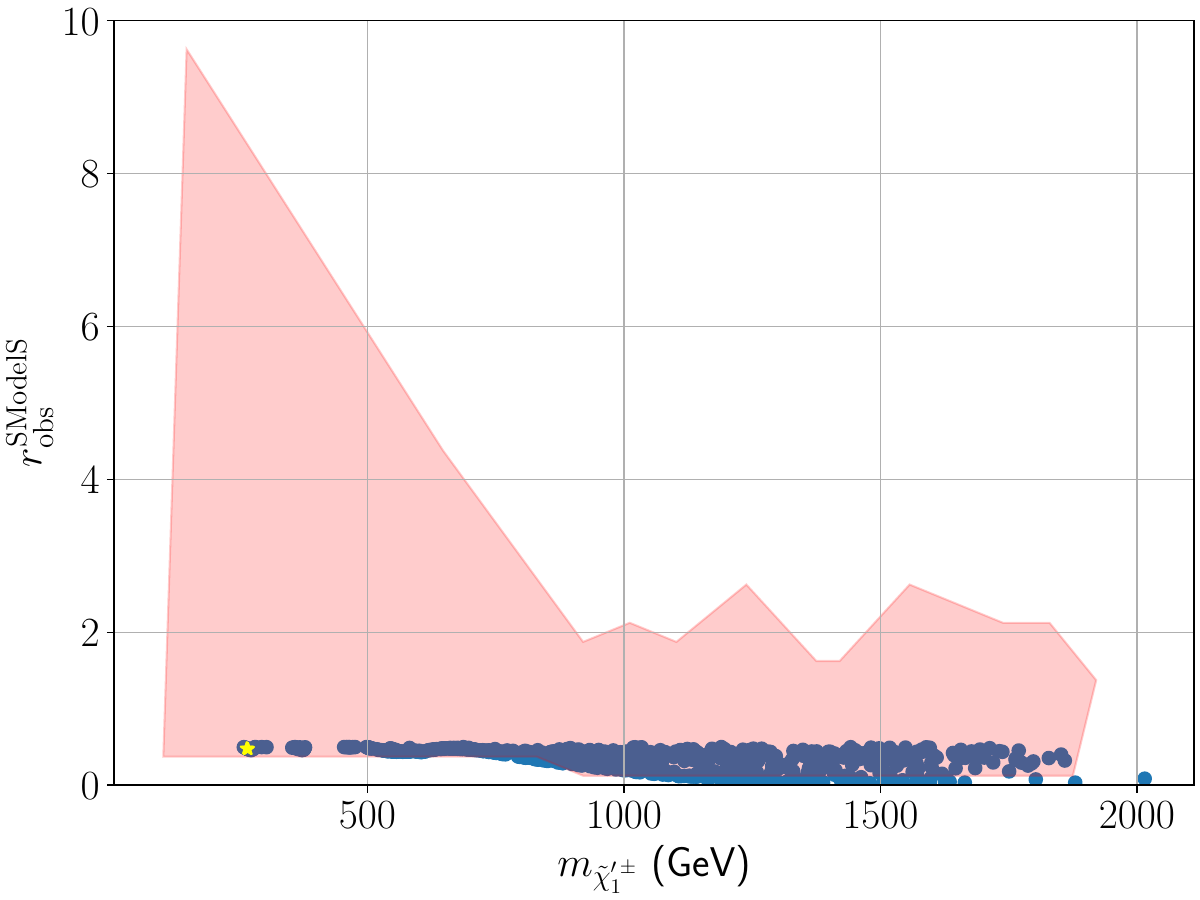} %
        \caption{}
    \end{subfigure}%
   \begin{subfigure}[b]{0.45\textwidth}
        \centering
        \includegraphics[width=\linewidth]{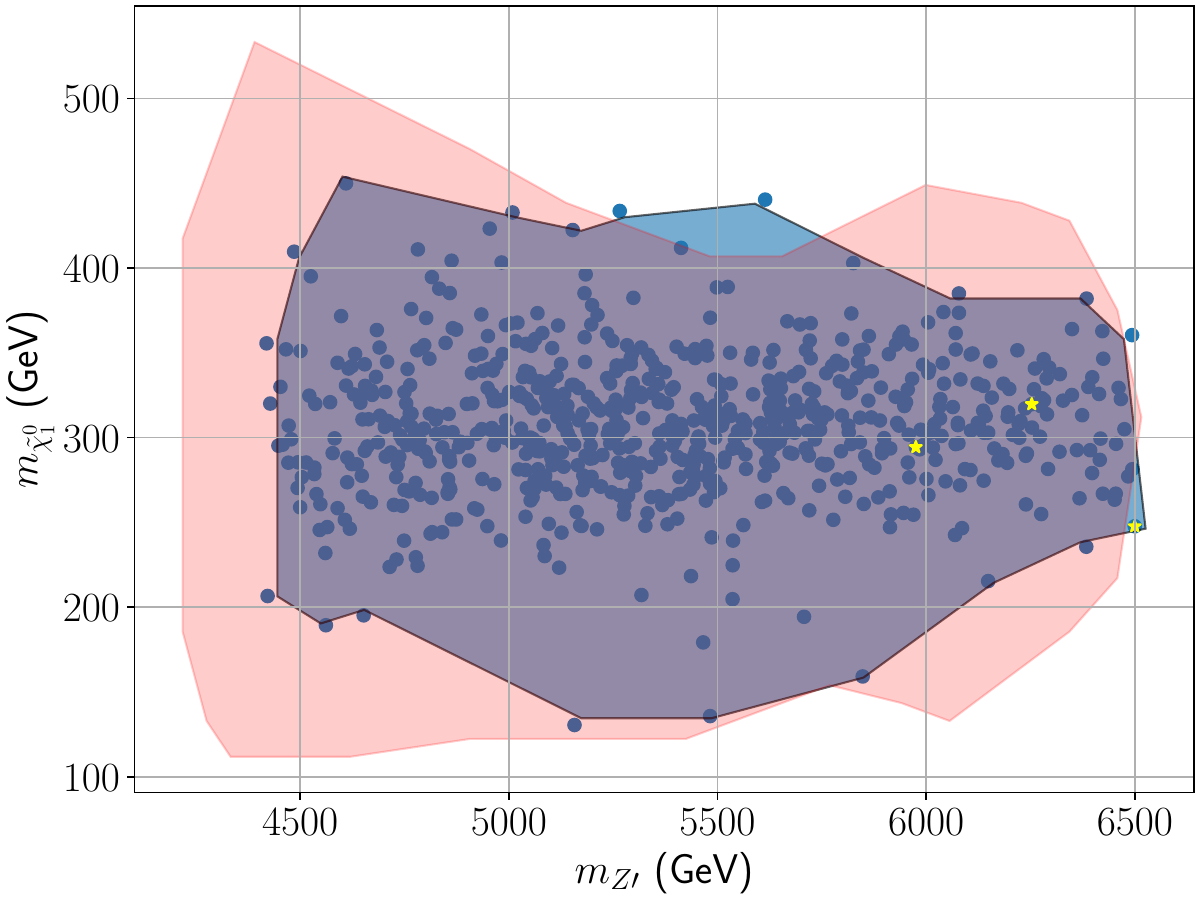}
        \caption{}
    \end{subfigure}
    \caption{\label{FIG:High3}Allowed parameter space for high-scale boundary conditions in terms of observables. (a) Shows the {\sc SModelS} signal-to-exclusion ratio against the lightest sterile chargino mass; (b) shows the $Z^\prime$ boson mass and (lack of) correlation with the lightest (visible) neutralino mass.}
\end{figure}

We collect together all parameter points obtained from this scan that produced a valid spectrum, and then process them, separating into ``good'' and ``bad'': the good points pass all constraints separately, including the more stringent requirement that the most sensitive ratio of cross-section to limit from {\sc SModelS} be less than $0.5$ instead of $1$ (to allow for NLO corrections and other uncertainties). We then plot the results in two-dimensional slices, showing the regions in which good points were found as green and those where bad points were found as red; the actual ``good'' points found are shown as blue dots, and we show three best-fit benchmark points as yellow stars. These plots are shown in figures \ref{FIG:Highm0}, \ref{FIG:HighmChaP} and \ref{FIG:High3}.

We see that enhancement to the mass of the W boson is possible for light sterile gaugino masses, so the maximum enhancement that can be obtained is cut off by the LHC lower limits on those particles. The strongest limits come from classic supersymmetry searches with missing energy and leptons in the final state \cite{ATLAS:2019lff,CMS:2020bfa} and from fully-hadronic final states\cite{CMS:2021beq}, although for some points the $\tilde{\chi}_2^{\prime\,+}$ and $\tilde{\chi}_3^{\prime\,0}$ become sufficiently close in mass that disappearing track searches apply and dominate. In general, the decay chains of the lighter sterile neutralino proceed via $\tilde{\chi}_2^0 \rightarrow \tilde{\chi}_1^{\prime\,0} + h/Z$ with roughly equal branching fractions; and $\tilde{\chi}_1^{\prime\,+} \rightarrow \tilde{\chi}_1^{\prime\,0} + W^+$ at nearly $100\%$: these decays release a lot of missing energy and can be excluded if the cross-section is large enough. On the other hand, the heavier sterile electroweakinos decay via cascades rather than directly to $\tilde{\chi}_1^{\prime\,0},$ (sometimes with a very small width), and this complicates the searches: examining the constraints with a full recast would be an interesting subject for future work.

Modifications to the W boson mass can also be enhanced by splitting the sterile gaugino masses, and this requires large couplings, in particular $\lambda_5.$ However, it is difficult to reconcile large values of $\lambda_5$ with running from the GUT scale. Hence, with high-scale boundary conditions, it is difficult to obtain a W boson mass compatible with the CDF result in the scenario we have considered here; but a value of $80.37$ GeV is plausible.

We also see from \ref{FIG:High3}(b) the robust lower bound on the $Z'$ boson mass in this model around $4.5$ TeV. This limit is dominated by the decays to muons, and at such large masses it is difficult to modify the relevant properties to change the limit; in general $Z'$ limits can be affected by opening up other decay channels or mixing with the $Z$, neither of which have much effect here. 

From our dataset we selected three best-fit benchmark points, whose parameters we show in table \ref{TAB:HighSpect}. Notably we can see that large $\tan \beta$ is favoured, along with $\lambda_5$ as large as possible to split the multiplet masses and enhance $M_W.$

\begin{table}[h!]\centering
\begin{tabular}{|c|c|c|c|}\hline\hline
 & HP1 & HP2 & HP3 \\ \hline\hline
$m_0$ (GeV)  & $3617.7211$  & $3759.2217$  & $3583.5602$ \\
$m_{1/2}$ (GeV)  & $1552.8587$  & $1978.6734$  & $1956.0013$ \\
$A_0$ (GeV)  & $-2245.5365$  & $-1452.2188$  & $-2465.4575$ \\
$\tan \beta$  & $38.3117$  & $22.1819$  & $33.8167$ \\
$v_S$ (GeV)  & $19991.4732$  & $18353.1278$  & $19275.0017$ \\
$\lambda$  & $0.0338$  & $0.0233$  & $0.0516$ \\
$b_0$ (GeV)  & $1225.2828$  & $1625.9454$  & $1118.0716$ \\
$\lambda^\prime$  & $0.0123$  & $0.0135$  & $0.0127$ \\
$\lambda_5$  & $0.4071$  & $0.4339$  & $0.4086$ \\
$\lambda_6$  & $-0.0062$  & $-0.0028$  & $0.0051$ \\
\hline\hline
$\mathrm{M}_W$ (GeV)  & $80.3678$  & $80.3692$  & $80.3667$ \\
$m_{Z\prime}$ (GeV)  & $6499.7388$  & $5975.1617$  & $6252.9409$ \\
$m_h$ (GeV)  & $123.3620$  & $123.4241$  & $124.7690$ \\
$m_{\tilde{\chi}^{\prime\,\pm}_1}$ (GeV)  & $266.8361$  & $266.6738$  & $264.6775$ \\
$m_{\tilde{\chi}^{\prime\,0}_2}$ (GeV)  & $283.1408$  & $284.0707$  & $281.1326$ \\
$m_{\tilde{\chi}^{0}_1}$ (GeV)  & $247.6976$  & $294.3537$  & $319.7145$ \\
$m_{\tilde{\chi}^{0}_2}$ (GeV)  & $445.2606$  & $-352.0849$  & $596.5003$ \\
$r_{\rm obs}^{\rm SModelS}$  & $0.4750$  & $0.4757$  & $0.4812$ \\
\hline\hline
\end{tabular}
\caption{\label{TAB:HighSpect} Inputs and spectrum of high-scale benchmark points.}
\end{table}

\subsection{Low-scale boundary conditions}

For low-scale boundary conditions, we take one point with high-energy input parameters $m_0 =3216$ GeV, $m_{1/2} = 1328$ GeV, $A_0 = -2383$ GeV and run to low energies, where we fix $\lambda_6 = 0.01, \mu = 4720$ GeV, $B_\mu = 5.97 \times 10^5\ (\GeV)^2, v_S = 13970\ \GeV. $ The input parameter card will be released along with the update for \BSMArt, so all the parameters at low energies can be consulted directly online.  

\begin{figure}[h!]
    \centering
   \begin{subfigure}[b]{0.45\textwidth}
        \centering
        \includegraphics[width=\linewidth]{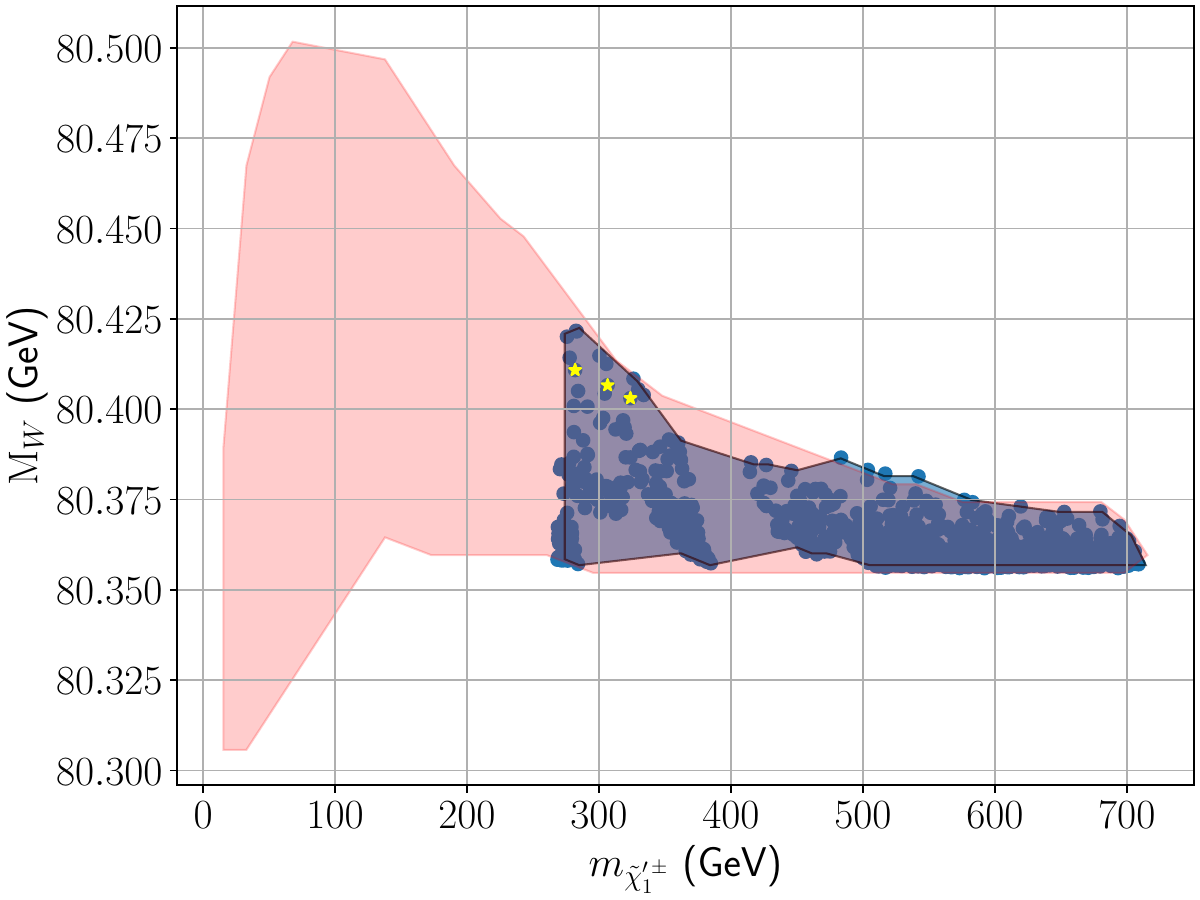} %
        \caption{}
    \end{subfigure}%
   \begin{subfigure}[b]{0.45\textwidth}
        \centering
        \includegraphics[width=\linewidth]{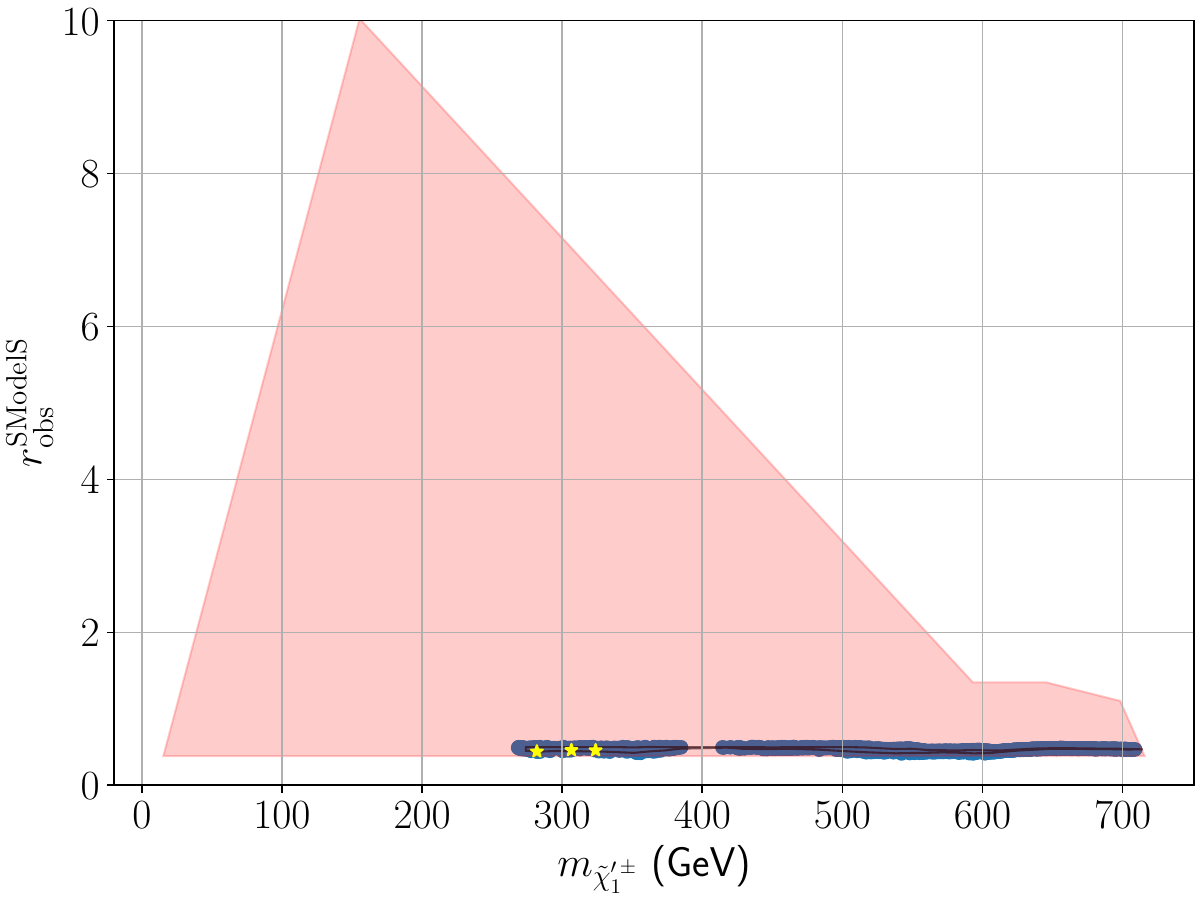}
        \caption{}
    \end{subfigure}
    \caption{\label{FIG:Low1}Allowed parameter space for low-scale boundary conditions. (a) Shows the dependence of the $W$ boson mass on the lightest sterile gaugino mass; (b) shows the {\sc SModelS} signal-to-exclusion ratio against the lightest sterile chargino mass.}
\end{figure}

\begin{figure}[h!]
    \centering
   \begin{subfigure}[b]{0.45\textwidth}
        \centering
        \includegraphics[width=\linewidth]{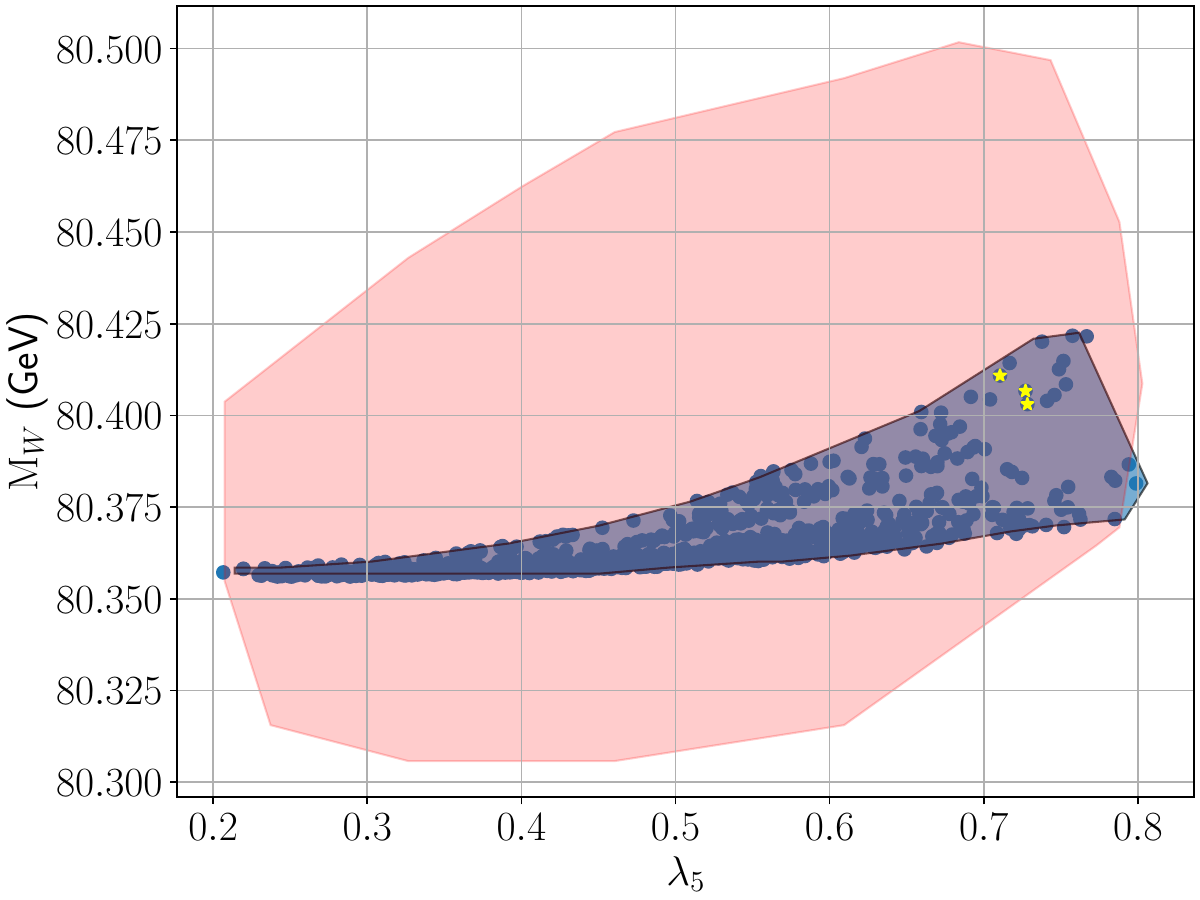} %
        \caption{}
    \end{subfigure}%
   \begin{subfigure}[b]{0.45\textwidth}
        \centering
        \includegraphics[width=\linewidth]{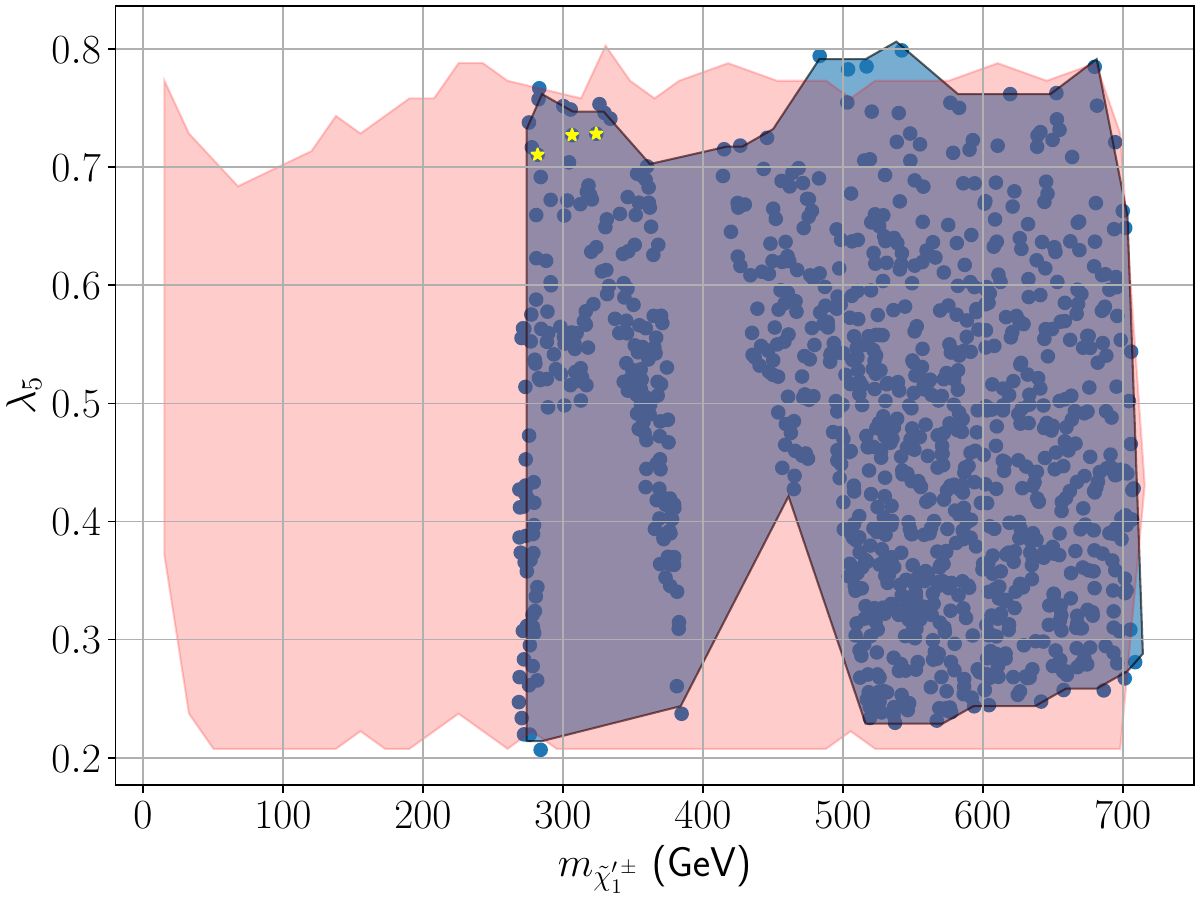}
        \caption{}
    \end{subfigure}
    \caption{\label{FIG:Low2}Allowed parameter space for low-scale boundary conditions. (a) shows the proportionality of the $W$ boson mass to the coupling $\lambda_5$; (b) shows the (lack of) correlation between the lightest sterile chargino mass and $\lambda_5.$ }
\end{figure}

We use the same MCMC likelihood function as before, but now scan over the following parameter ranges:
\begin{gather}
\tan \beta \in [2,10],\ \lambda^\prime \in [-0.07,0.07],\ \lambda_5 \in [0.2,0.8].
\end{gather}
The results are shown in figure \ref{FIG:Low1} and \ref{FIG:Low2}. We see that enhancement to the mass of the W boson is possible for large $\lambda_5$ (although large values are difficult to reconcile with running from the GUT scale) and that very large values are only possible for light sterile gaugino masses, so the maximum enhancement that can be obtained is cut off by the LHC lower limits on those particles.

As for the high-scale conditions, we also selected three benchmark points with the best scores, shown in table \ref{TAB:LowSpect}; these are all very similar in nature, and near the boundary of acceptable values for the LHC limits from {\sc SModelS}. 
\begin{table}[h!]\centering
\begin{tabular}{|c|c|c|c|}\hline\hline
 & LP1 & LP2 & LP3 \\ \hline\hline
$\tan \beta$  & $2.5823$  & $2.3943$  & $2.4802$ \\
$\lambda^\prime$  & $0.0278$  & $0.0302$  & $0.0319$ \\
$\lambda_5$  & $0.7106$  & $0.7271$  & $0.7283$ \\
\hline\hline
$\mathrm{M}_W$ (GeV)  & $80.4109$  & $80.4067$  & $80.4031$ \\
$m_{Z\prime}$ (GeV)  & $4514.4299$  & $4514.4309$  & $4514.4332$ \\
$m_h$ (GeV)  & $126.4964$  & $124.9597$  & $125.9505$ \\
$m_{\tilde{\chi}^{\prime\,\pm}_1}$ (GeV)  & $281.9345$  & $306.5685$  & $323.8345$ \\
$m_{\tilde{\chi}^{\prime\,0}_2}$ (GeV)  & $284.3340$  & $309.1594$  & $326.5368$ \\
$m_{\tilde{\chi}^{0}_1}$ (GeV)  & $203.8493$  & $203.5185$  & $203.6725$ \\
$m_{\tilde{\chi}^{0}_2}$ (GeV)  & $366.0128$  & $365.9468$  & $366.0067$ \\
$r_{\rm obs}^{\rm SModelS}$  & $0.4417$  & $0.4593$  & $0.4576$ \\
\hline\hline
\end{tabular}
\caption{\label{TAB:LowSpect} Spectra for low-scale benchmark points.}
\end{table}

\section{Conclusions}
\label{SEC:CONCLUSIONS}

We have developed a new phenomenological model describing an $E6$ extension of the Supersymmetric Standard Model involving approximate symmetries. We derived approximate expressions for the mass matrices to understand the phenomenology of the new ``sterile'' sectors. We introduced a framework for applying all relevant low-energy constraints, and studied the properties of the model, first starting from high-scale boundary conditions, and then from low-scale conditions. 

We found a robust lower limit on the $Z'$ mass, and found the conditions for the sterile gaugino sector to enhance the $W$ boson mass. We found that larger enhancements are in tension with LHC searches and also with the absence of Landau poles at the GUT scale, but the CDF result (or a world average) could in principle be accommodated.

We expect that future LHC searches should probe the relevant part of the parameter space. Alternatively, it would be interesting to explore the impact of powerful recent searches such as \cite{ATLAS:2021yqv,ATLAS-CONF-2023-048} for which no recast is currently available, or apply full recasts of other searches to our benchmarks to test sensitivity. These are beyond the scope of this paper.

It would also be very interesting in future work to explore other variants of this model; with more effort a version with multiple Higgs expectation values could be implemented, for example; or a version without the approximate global symmetries; or even a version with an explicit breaking of our $U(1)_H.$ Moreover, due to the complexity of the model it would be beneficial to apply (yet) more sophisticated scanning algorithms to better explore its parameter space.

\section*{Acknowledgments}

This work was supported by a Royal Society Exhange grant IES\textbackslash R1\textbackslash 221199, ``Phenomenological studies of string vacua''. AEF Thanks the theory group at CERN for support and hospitality.
M.~D.~G. acknowledges support from the ANR via grant \mbox{``DMwithLLPatLHC''}, (ANR-21-CE31-0013). 
We thank Junichiro Kawamura for discussions about bounds on vector-like leptons, Asesh Datta for correspondence about BSMArt, and 
Marco Guzzi for collaboration at the initial stage of the project. 

\appendix

\section{New additions in \BSMArt v1.2 and \SARAH v4.15.2}
\label{APP:BSMArt}

For this study, it was necessary to add the following tools to \BSMArt, and the associated installation scripts. The new code is available at \url{https://goodsell.pages.in2p3.fr/bsmart/}.

\subsection{{\tt QNUMBERS} blocks from \SARAH and \BSMArt}

 The latest version of \SARAH, available at \url{https://sarah.hepforge.org}, will create a text file containing {\tt QNUMBERS} blocks, with the quantum numbers of all BSM particles, for use by various codes but notably here \SModelS. The same routines are included in \BSMArt so that the blocks can be created using older versions of \SARAH. 
 
 Since \SModelS requires information about ``odd particles,'' these are determined automatically for supersymmetric models from R-parity, but -- and especially for the model considered in this paper -- it is also possible to specify them by giving a list in the {\tt SPheno.m}:
 \begin{lstlisting}[language=Mathematica,basicstyle=\scriptsize,numbers=none,caption=E6AFH/SPheno.m]
     ExtraOddParticles={ChiI,ChiP,SHI0,SHIp};
 \end{lstlisting}

\subsection{Cross-sections from \MicrOMEGAs}

Two new main files for \MicrOMEGAs to use \SARAH models are provided: {\tt MicrOmegas\_v5.3\_BSMArt\_xsections.cpp} and {\tt MicrOmegas\_v5.3\_BSMArt\_onlyxsections.cpp}. These compute cross-sections at 8 TeV and 13 TeV for pairs of particles, as well as extracting information about any $Z'$ gauge bosons required for {\sc ZPEED}, and computing the resonant production cross-section for {\sc Z' Explorer} (see below). The first of the files is relevant for dark matter studies, as it will also compute all of the relevant cross-sections when it is called with the {\tt --xsections} flag; the second is relevant for models such as the one considered in this paper, since it will not compute the relic density (the model does not have a good dark matter candidate). 

Example code in the scan {\tt json} file to launch this would be:
\begin{lstlisting}[language=python,basicstyle=\scriptsize,numbers=none]
"Codes": {
   "MicrOmegas": {
	    "Command": "/path/to/micromegas/SARAH_E6AFH/MicrOmegas_v5.3_BSMArt_onlyxsections",
	    "Cross-sections":"True",
	    "InputFile": "SPheno.spc.E6AFH",
	    "Run":"True"
	},
}
\end{lstlisting}

\subsection{{\sc ZPEED}}

This is a python code to compute $Z'$ limits in the dilepton channel, including effects from interference. It is then imported as part of the \BSMArt tool, requiring only that a variable {\tt Path} be specified in the tool settings. During runtime, it will read the required information from the spectrum file which must contain a block {\tt ZPEEDIN}. This comprises the 
\begin{lstlisting}[language=SLHA,basicstyle=\scriptsize,numbers=none]
BLOCK ZPEEDIN #
 1   Zppdg # Z' pdg number
 2   gdv  # Vector coupling of down-type quarks
 3   gda  # Axial vector coupling of down-type quarks
 4   guv  # Vector coupling of up-type quarks
 5   gua  # Axial vector coupling of up-type quarks
 6   gev  # Vector coupling of the electron
 7   gea  # Axial vector coupling of the electron
 8   gmuv  # Vector coupling of the muon
 9   gmua  # Axial vector coupling of the muon
\end{lstlisting}
In addition, the width of the $Z'$ gauge boson is read from the {\tt DECAY} blocks. These inputs will be automatically written to the spectrum file by the above \MicrOMEGAs main files, which will even detect $Z'$ bosons automatically, so there is no need for the user to write these blocks themselves! In fact, the case of multiple $Z'$ bosons can also be handled by information in the {\tt ZPEEDIN} block advanced by ten (so entries 11--19 for the second, 21--29 for the third etc).

The outputs are written/stored in the {ZPEED} block:
\begin{lstlisting}[language=SLHA,basicstyle=\scriptsize,numbers=none]
BLOCK ZPEED #
 1 %10.6E # -2 log L
 2 %10.6E # -2 Delta log L
 3 %10.6E # Best CLs
\end{lstlisting}

Example settings to use this tool are:
\begin{lstlisting}[language=python,basicstyle=\scriptsize,numbers=none]
"Codes": {
   "ZPEED":{
	    "Path": "/path/to/ZPEED/ZPEED-master",
	    "Run": "True",
	    "Observables": {
		"loglike": {"SLHA": ["ZPEED",[1]], "SCALING": "MINUSEXPUSER"},
		"CLs" : { "SLHA":["ZPEED",[3]], "SCALING": "OFF", "MEAN": 0.05, "VARIANCE": 0.01}
		}
	}
}
\end{lstlisting}

\subsection{{\sc Z' Explorer}}

{\sc Z' Explorer} is a c code which computes limits based on interpolated cross-sections and scraped limits. The implementation in \BSMArt is actually a pythonisation: the data tables are bundled with the code, except the cross-sections are read from the {\tt SLHA} file rather than interpolated; this information is computed in the \MicrOMEGAs code above. The results are stored in the output {\tt SLHA} file in a block {\tt ZPRIME}:
\begin{lstlisting}[language=SLHA,basicstyle=\scriptsize,numbers=none]
BLOCK ZPRIME #
 1 %10.6E # Best Limit
 2 %d # pdg for best limit CLs
\end{lstlisting}
A point is considered ruled out if the best limit is greater than unity. 

The only setting needed in the \BSMArt code is the pdg code(s) of any $Z'$ boson(s), given as a list:
\begin{lstlisting}[language=python,basicstyle=\scriptsize,numbers=none]
"Codes": {
    "Zprime": {
        "PDGs": [31],
        "Run": True,
        "Observables": {"zplim": {"SLHA":["ZPRIME",[1]], "SCALING":"OFF"}}
    }
}
\end{lstlisting}

\subsection{{\sc SModelS}}

The install script for \BSMArt can now install \SModelS, and set up the scan. The only required input paramter is the path to the parameter file, but there is also the option to specify a path for the {\tt QNUMBERS} file and the location of the cache (so that the large \SModelS cache can be stored inside the source directory rather than in the user's home directory). The cross-sections must be included in the input {\tt SLHA} file, but again these can be computed by \MicrOMEGAs. 

The output of the code is:
\begin{lstlisting}[language=SLHA,basicstyle=\scriptsize,numbers=none]
BLOCK SMODELS #
 1 %10.6E # Best R ratio (observed)
 2 %10.6E # r - 1: > 0 if excluded, < 0 if allowed
\end{lstlisting}
The R ratio is the ratio of the cross-section to best limit, as used in the text. In the comments of the output it will also print the analysis that gives the best limit. 

Thus the json code to tell \BSMArt to run \SModelS could look like:
\begin{lstlisting}[language=python,basicstyle=\scriptsize,numbers=none]
"Codes": {
    "Path": "/path/to/smodels-main/",
		"Cache": "/path/to/smodels-main/CACHE",
		"QNUMBERS file": "QNUMBERS_E6AFH.slha",
		"Parameter file": "/path/to/parameters.ini",
		"Observables": {"SModelSr": {"SLHA":["SMODELS",[1]], "SCALING": "UPPER", "MEAN": 1.0, "VARIANCE": 0.2}, "SModelSrm1": {"SLHA":["SMODELS",[2]], "SCALING":"OFF"}},
             	"Run": "True"
}
\end{lstlisting}

\section{Lepton number model}
\label{APP:LeptonNo}

\begin{table}[h]
\begin{center}
{\small Additional chiral and gauge multiplet fields in the {\tt E6AFL} model}\\
\begin{tabular}{|c|c|c|c|c|c|c|c|}
\hline
  Superfield               & Generations  &Scalars                 & Fermions & Vectors &  & $U(1)_{Z'}$ & $N_L$\\ \hline
$\mathbf{W_{Z',\alpha}}$ & 1& & $\tilde{B}'$ & $B^\prime_\mu$ & (\textbf{1}, \textbf{1}, 0 ) & 0 & 0 \\ \hline
  $\mathbf{S}_0$&1& $S_0$ & $F_{S}$   & & (\textbf{1},\textbf{1},0) & -2 & 0 \\
  $\mathbf{S}_P$&1& $S_P$ & $F_{SP}$   & & (\textbf{1},\textbf{1},0) & -2 & 1\\
  $\mathbf{S}_M$&1& $S_M$ & $F_{SM}$   & & (\textbf{1},\textbf{1},0) & -2 & -1\\
    $\mathbf{\phi}$&1& $ \Phi$ & $F_{\phi}$   & & (\textbf{1},\textbf{1},0) & 1 & 1 \\
$\mathbf{\tilde{\phi}} $&1& $ \tilde{\Phi}$ & $F_{\tilde{\phi}}$   & & (\textbf{1},\textbf{1},0) & -1 & -1 \\
  $\mathbf{X_d}$&1 & $(X_d^0 , X_d^-)$ & $(\tilde{X}_d^0 , \tilde{X}_d^-)$ & & (\textbf{1}, \textbf{2}, -1/2) & $-4/5$ & 1\\
  $\mathbf{X_u}$&1  & $(X_u^+ , X_u^0)$ & $(\tilde{X}_u^+ , \tilde{X}_u^0)$ & & (\textbf{1}, \textbf{2}, 1/2) & $4/5$ & -1\\
  $\mathbf{H_{1 I=\{1,2\}}}$ &2 & $(H_{1I}^0 , H_{1I}^-)$ & $(\tilde{H}_{1I}^0 , \tilde{H}_{1I}^-)$ & & (\textbf{1}, \textbf{2}, -1/2) & $6/5$ & 1\\
  $\mathbf{H_{2 I=\{1,2\}}}$ &2 & $(H_{2I}^+ , H_{2I}^0)$ & $(\tilde{H}_{2I}^+ , \tilde{H}_{2I}^0)$ & & (\textbf{1}, \textbf{2}, 1/2) & $4/5$ & -1 \\
  $\mathbf{D_x}$&1 & $ D_x$ & $F_{D_x} $ & & (\textbf{3}, \textbf{1}, -1/3) & $4/5$ & 0 \\
   $\mathbf{\tilde{D}_x}$&1 & $ \tilde{D}_x$ & $F_{\tilde{D}_x} $ & & ($\mathbf{\ov{3}}$, \textbf{1}, 1/3) & $6/5$ & 0\\
\hline
\end{tabular}
\caption{\label{TAB:LNfields}Field content of the lepton model supplementary to the MSSM fields.} 
\end{center}
\end{table}

In this appendix we give an alternative formulation of the theory, where the new fields are all assinged lepton number; lepton number should be explicitly broken by a neutrino mass, but for the purposes of computing collider constraints this can be neglected. We give the quantum numbers of the MSSM and new fields under this symmetry in table \ref{TAB:LNfields}. 

The superpotential of the this theory is:
\begin{align}
  W =& Y_u u q H_u - Y_d d q H_d - Y_e e l  H_d  + \mu X_u X_d + \lambda s H_u H_d + \lambda_{12}^{IJ} s H_{2I} \cdot H_{1J} \nn\\
     &+ \kappa s D_x D_{\ov{x}} + g_{QL} D_x q q + g_{QR} D_{\ov{x}} d u + \mu^\prime \phi \tilde{\phi}\nn\\
  &+ \lambda_5 S_M H_u H_{1I} + \lambda_6 S_P H_{2I} H_d +Y_x e X_d H_d + Y_\ell^\prime e H_{1I} H_d.
\end{align}
The inclusion of the mixing term $Y_\ell^\prime$ means that the heavy fermions can decay to leptons, and this exposes them to heavy vector-like lepton searches at the LHC. 
The fields $H_u, H_d$ obtain an expectation value, as does the singlet $S_0 = \frac{1}{\sqrt{2}} (v_S + S_R + i S_I)$, with $v_S \gg v$. The resulting mixings of the fields are:

\begin{center}
\begin{tabular}{|c|c|c|}
\hline
  Multiplet     & Number  of fields  & Gauge eigenstates    \\ \hline
  Neutralinos & 6 & $ \tilde{B}, \tilde{W}^0, \tilde{H_d^0}, \tilde{H_u^0}, F_{S}, \tilde{B}^\prime$ \\
  Charginos & 2 & $(\tilde{W}^+, \tilde{H}_d^+),(\tilde{W}^-, \tilde{H}_u^- )$ \\
  Charged leptons & 6 & $(e_{i,L}, \tilde{X}_d^-, \tilde{H}_{1I}^-), (\tilde{e}_{i,R}, \tilde{X}_u^+, \tilde{H}_{2I}^+)$ \\
  Heavy (dirac) neutrinos & 5 & $(\tilde{H}_{1I}^0,\tilde{X}_d^0, F_{S_P}, F_{\phi}), (\tilde{H}_{2I}^0,\tilde{X}_u^0, F_{S_M}, F_{\tilde{\phi}})$ \\
  Selectrons & 12 & $e_{i,L}, (e_{i,R})^*, H_{1I}^-, (H_{2I}^+)^*, X_d^-, (X_d^+)^*$ \\
  Sneutrinos & 11 & $\tilde{\nu}_L, H_{1I}^0, (H_{2I}^0)^*, X_d^0, (X_u^0)^*, \Phi, \tilde{\Phi}^*$ \\
\hline
\end{tabular}
\end{center}


\bibliographystyle{JHEP}
\bibliography{references}

\end{document}